\documentclass[
reprint,
superscriptaddress,
longbibliography,
bibnotes,
amsmath,amssymb,
aps,
prb,
]{revtex4-2}
\usepackage[utf8]{inputenc}
\usepackage[english]{babel}
\usepackage[plainpages = false, pdfpagelabels, 
                 bookmarks,
                 bookmarksopen = true,
                 bookmarksnumbered = true,
                 breaklinks = true,
                 linktocpage,
                 colorlinks = true,
                 linkcolor = blue,
                 urlcolor  = blue,
                 citecolor = blue,
                 anchorcolor = green,
                 hyperindex = true,
                 hyperfigures
                 ]{hyperref} 
\usepackage{enumerate}
\usepackage{tabularx}
\usepackage{makecell, multirow}
\usepackage{tikz}
\usetikzlibrary{patterns}
\usepackage{slashed}
\usepackage{physics}
\usepackage{verbatim}
\usepackage{svg}
\usepackage{mathtools}
\usepackage{cancel}
\usepackage{bbold}
\usepackage{bm}
\usepackage{soul}
\usepackage{booktabs}
\usepackage[caption=false,singlelinecheck=off]{subfig}
\usepackage{blindtext}
\usepackage{bbm}
\usepackage{graphicx}
\usepackage{dcolumn}
\usepackage{bm}
\usepackage[normalem]{ulem}
\usepackage{xcolor}

\usepackage[percent]{overpic}
\usepackage{subfig}
\begin{document}
\title{Extracting the anyon charge from shot noise in complex fractional quantum Hall edges
}

\author{Jinhong Park}
\affiliation{\mbox{Department of Physics, Konkuk University, Seoul 05029, Republic of Korea}}

\author{Christian Sp{\r a}nsl{\"a}tt}
\affiliation{\mbox{Department of Engineering and Physics, Karlstad University, Karlstad, Sweden}}

\begin{abstract}
Shot-noise measurements with a quantum point contact provide a direct probe of the quasiparticle charge in Laughlin fractional quantum Hall (FQH) states. However, the interpretation of such measurements becomes more subtle for multichannel edges, where the point-contact tunneling noise coexists with noise generated by intermode equilibration and heating. We develop a general framework that incorporates both these contributions and apply it to the paradigmatic $\nu=2/3$ FQH edge. We derive expressions for the drain auto- and cross-correlations in a four-terminal QPC geometry and show that equilibration-induced heating affects the auto- and cross-correlations in parametrically different way than the intrinsic partition noise. This distinction permits us to propose a specific combination of the two correlations which strongly suppresses the heating contribution and fully recovers the tunneling quasiparticle charge. Our results thereby provide a systematic framework for interpreting shot-noise measurements in complex FQH edges and can be further extended to more general counterpropagating edge structures.
\end{abstract}

\maketitle
\date{\today}

\section{Introduction}
Shot noise, or partition noise, originates from the granularity of electric charge and has developed into a powerful probe of elementary excitations in mesoscopic conductors~\cite{Blanter2000,Kobayashi2021}. In particular, the effective charge inferred from shot noise can reveal the nature of the underlying transport carriers. This is demonstrated, for example, by measuring the charge-$2e$ transfer associated with Andreev processes in superconducting hybrid structures~\cite{Jehl2000} and by measurements of the charge carried by Bogoliubov quasiparticles~\cite{Ronen2016}. One of the most celebrated applications of shot noise is in the fractional quantum Hall (FQH) effect, where shot-noise measurements with a quantum point contact (QPC) provided direct evidence for the fractional charge of anyons in the Laughlin states~\cite{dePicciotto1997,Saminadayar1997,Heiblum2020edge}.

In this work, we study FQH states where the interpretation of shot noise becomes considerably more subtle. When the edge hosts multiple, counterpropagating edge modes, intermode equilibration---the tendency of edge channels to reach the same local electrochemical potential or temperature---plays a central role in the charge and energy transport. The impact of interactions and disorder on the equilibration process and its impact on electrical and thermal transport have been studied in detail, particularly for the paradigmatic $\nu=2/3$ edge~\cite{Protopopov_transport_2_3_2017,Nosiglia2018,Spanslatt2021}. Importantly, charge and heat transport can exhibit very different equilibration behavior~\cite{kane_quantized_1997}, making thermal transport complementary to electrical transport for probing the underlying edge structure. This has been  experimentally demonstrated with measurements of quantized heat flow in Abelian FQH states~\cite{Banerjee2017, Srivastav_Universal_quantized_2019, Srivastav2021May,Srivastav2022Sep,Breton2022}, and subsequently played a key role in probing the non-Abelian $\nu=5/2$ state~\cite{Banerjee_observation_2018,Dutta_Isolated_2022,Paul2024Aug}. 

An important aspect of edge equilibration is that the noise it generates can by itself carry information about the underlying topological order. For multichannel Abelian FQH edges, it was shown that equilibration-induced noise exhibits characteristic length dependencies determined by the direction of heat flow~\cite{Park_Noise_2019,Spanslatt_Noise_2019}. This framework was further extended to QPC geometries, where the noise accompanying quantized conductance plateaus was shown to exhibit characteristic scaling with the QPC and device dimensions~\cite{Spanslatt_condplateau}; see also a closely related recent work based on an equilibration model~\cite{Glattli2025}. The equilibration picture has received experimental support from observations of the coherent-to-incoherent crossover~\cite{Hashisaka2023} and from measurements of upstream noise in hole-conjugate FQH states, revealing a pronounced separation between charge and thermal equilibration scales~\cite{Melcer_Absent_2022,Kumar2022Jan}. The same idea was subsequently extended to the non-Abelian $\nu=5/2$ state, where the length dependence of noise, together with thermal transport, was proposed as a means of distinguishing competing topological orders~\cite{Park_noise_2020,Yutushui_Identifying_2023,Hein_Thermal_2023}, and it was supported by partition-noise measurements at engineered $\nu=5/2$ interfaces providing further information on the underlying edge structure~\cite{Dutta_novel_2022}. The mechanisms for the formation of conductance plateaus and their noise signatures at $\nu = 5/2$ states were explored in Ref.~\cite{Park2024Jun,Manna2025}.

Despite these advances, the extraction of quasiparticle charge from QPC shot noise has not been established in multichannel FQH edges, in sharp contrast to the Laughlin states. Indeed, anomalous effective charges and finite noise on conductance plateaus have long been observed in hole-conjugate states~\cite{Bid2009,Sabo2017,Bhattacharyya2019Jun,Biswas2022Dec}, highlighting the nontrivial role of the edge structure and upstream propagation in the interpretation of QPC-generated shot noise. For such edges, the intrinsic partition noise generated at the QPC coexists with additional noise arising from intermode equilibration and heating, making it essential to distinguish the two contributions. A recent experiment in Ref.~\cite{De_HOM_2025} at $\nu=2/3$ partially addressed this issue by comparing auto- and cross-correlations, but the heating contribution was treated phenomenologically, leaving a microscopic understanding of how equilibration-induced noise enters the measured noise-correlations unsettled. It is therefore desirable with a theory that includes a joint description of QPC partitioning and the noise arising from edge equilibration.

Here, we present a general framework that achieves precisely this. As a paradigmatic example of a complex multichannel edge, we focus on the $\nu = 2/3$ FQH state. We derive both auto- and cross-correlations and show that equilibration-induced heating affects these quantities in parametrically differently way than the intrinsic partition noise at the QPC. This distinction enables us to define a suitable combination of the two correlations for which the heating contribution is substantially suppressed. This allows us to isolate the quasiparticle charge associated with the QPC partitioning process. Our approach provides a systematic route to interpret shot-noise measurements in complex FQH edges and can further be extended to more general counterpropagating edge structures, potentially including non-Abelian edges. 

The remainder of the paper is organized as follows. In Sec.~\ref{sec:setupmodel}, we introduce the setup and present the two basic ingredients of our framework: a single, equilibrating edge segment and a QPC geometry connecting four such segments. In Sec.~\ref{sec:linejunction}, we analyze charge and heat equilibration along the single edge segment, determine its voltage and temperature profiles, and derive the resulting noise due to equilibration. In Sec.~\ref{sec:fourterminalQPC}, we combine these results with QPC partitioning and derive expressions for the auto- and cross-correlations in the full four-terminal geometry. Finally, in Sec.~\ref{sec:Conclusion}, we summarize the work and discuss future directions.

Throughout this paper, we denote the conductance quantum of a single, chiral integer edge channel by
$G_0\equiv e^2/h$, where $e$ is the elementary charge and $h$ is the Planck constant.

\section{Setup and model}
\label{sec:setupmodel}

\begin{figure}[t]
 \centering
 \includegraphics[width=\columnwidth]{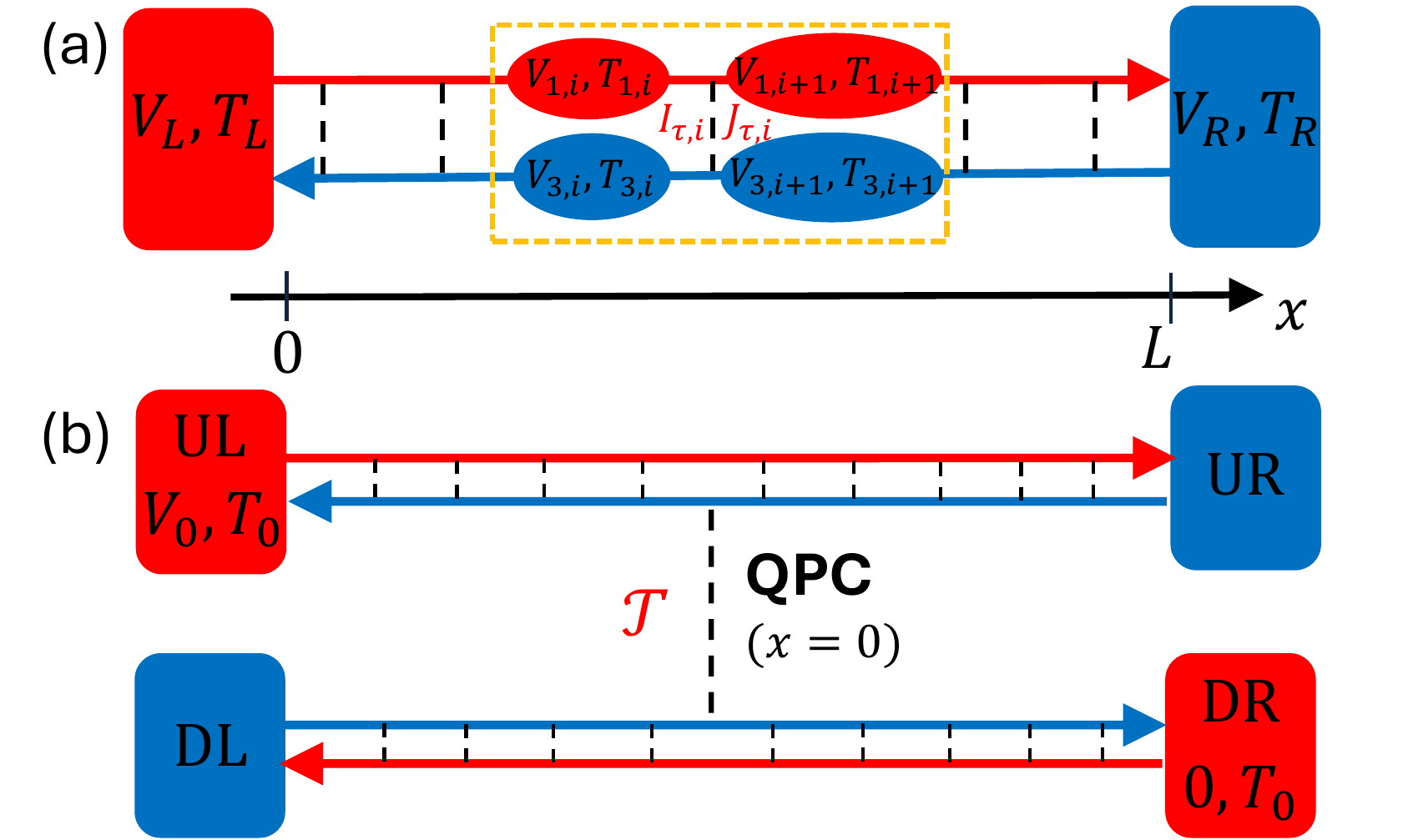}
 \caption{ (a) Model of an equilibrating $\nu=2/3$ line junction of length $L$. 
The red and blue lines represent the downstream $\nu=1$ integer channel and the upstream $\nu=1/3$ fractional channel, respectively. 
The line junction is divided into small cells, where the two counter-propagating channels exchange charge and energy through weak equilibration links. The highlighted yellow box illustrates the $i$th cell and the local voltages and temperatures on its two sides, with the corresponding tunneling charge and energy currents $I_{\tau,i}$ and $J_{\tau,i}$. 
The left and right boundary contacts are characterized by $(V_L,T_L)$ and $(V_R,T_R)$, respectively.
(b) Four-terminal QPC geometry. Each of the four arms is described by the equilibrating line-junction model shown in (a). A dc voltage $V_0$ is applied to the UL contact, while the DR contact is grounded. All four contacts are maintained at a base temperature $T_0$, and currents and their fluctuations are measured at the drain contacts UR and DL.
At the QPC located at $x = 0 $, the integer channel passes through, whereas the fractional channel is partially transmitted with inter-edge transmission probability $\mathcal{T}$.
}
 \label{fig:setup}
 \end{figure}

Our theoretical description consists of two ingredients, illustrated separately in Fig.~\ref{fig:setup}. We first introduce an equilibrating $\nu=2/3$ edge segment, which serves as the basic building block of our model [see Fig.~\ref{fig:setup}\textcolor{blue}{(a)}] and is analyzed in Sec.~\ref{sec:linejunction}. We then construct the four-terminal geometry by connecting four such edge segments through a single QPC [see Fig.~\ref{fig:setup}\textcolor{blue}{(b)}], as discussed in Sec.~\ref{sec:fourterminalQPC}. This separation allows us to distinguish equilibration processes occurring along the edge segments from partitioning processes occurring locally at the QPC.

\subsection{Equilibrating line junction}
\label{sec:linejunctionintro}

The basic building block of our model is a $\nu=2/3$ edge segment consisting of two counter-propagating bare channels: a downstream $\nu=1$ integer channel and an upstream $\nu=1/3$ fractional channel. Charge and energy are exchanged between these channels as they propagate along the edge. Charge equilibration is characterized by a length scale $\ell_{\rm ch}$, while thermal equilibration is characterized by a generally different length scale $\ell_{\rm th}$. Importantly, these two length scales can differ by orders of magnitude. Experiments on hole-conjugate FQH edges have demonstrated efficient charge equilibration while heat remains only weakly equilibrated over much longer distances, realizing the regime $\ell_{\rm ch} \ll \ell_{\rm th}$~\cite{Srivastav2021May,Melcer_Absent_2022, LeBreton2022}. Motivated by these observations, throughout this work we focus on the regime of strong charge equilibration, $\ell_{\rm ch}\ll L$, while we allow general ratios $\ell_{\rm th}/L$ in order to describe different regimes of thermal equilibration.

The coarse-grained description of such equilibration is illustrated in Fig.~\ref{fig:setup}\textcolor{blue}{(a)}. A line junction of length $L$ is divided into short cells of length $a$. In each cell, depicted as yellow, dashed box, the two counter-propagating bare channels are coupled by a weak tunneling link representing inter-mode scattering along the edge. The local states of the integer and fractional channels are characterized by effective voltages, $V_1$ and $V_3$, and effective temperatures, $T_1$ and $T_3$, respectively~\cite{SenAgarwal2008,Nosiglia2018}. Charge equilibration is described by the spatial variation of the local voltages, while thermal equilibration is described by the spatial variation of the local temperatures.

The same discrete description can also be used for current fluctuations. At sufficiently low frequencies, charge accumulation within an individual cell can be neglected, so that both the average current and its fluctuations are locally conserved. In addition to the fluctuations induced by the local voltage difference, each equilibration link produces an intrinsic tunneling-current fluctuation. The above description allows us to determine how the noise generated at different positions along a line junction propagates toward its outer boundaries. In Sec.~\ref{sec:linejunction}, we derive the continuum equations for the voltage, temperature, and voltage-fluctuation profiles by taking the continuum limit of this discrete model.

\subsection{Four-terminal QPC geometry}
The experimental geometry considered in this work is shown in Fig.~\ref{fig:setup}\textcolor{blue}{(b)}. It consists of two $\nu=2/3$ fractional quantum Hall edges connected by a single QPC, with the four edge segments attached to the contacts UL, UR, DL, and DR. Each of these four arms is modeled by the equilibrating line junction described in Sec.~\ref{sec:linejunctionintro}. A dc bias voltage $V_0$ is applied to the upper-left contact UL, while the lower-right contact DR is grounded. The output currents are measured at the two drain contacts UR and DL.

The QPC is treated as a single scattering element distinct from the equilibration processes occurring along the arms. We assume that the integer channel passes through the QPC without inter-edge tunneling, whereas the upstream $\nu=1/3$ channel is partially transmitted across the QPC with inter-edge transmission probability $\mathcal{T}$. Partitioning at the QPC provides an additional source of current fluctuations, which is treated separately from the intrinsic fluctuations generated by equilibration along the four arms.

The observables of our main interest are the low-frequency auto-current correlations at the two drains, $S_{\rm UR,UR}$ and $S_{\rm DL,DL}$, and the corresponding cross-current correlation $S_{\rm UR,DL}$, as will be discussed in detail in Sec.~\ref{subsec:autocrosscorrelation}. By first characterizing the propagation of fluctuations through an individual line junction and subsequently connecting the four line junctions through the QPC, we can identify separately the contributions originating from the arms and from QPC partitioning. We will show that these different sources of noise contribute to the auto- and cross-correlations in very different ways.

\section{Line junction}
\label{sec:linejunction}

We first consider a single equilibrating line junction of length $L$, parametrized by the coordinate $x$ with $0\leq x\leq L$, as illustrated in Fig.~\ref{fig:setup}\textcolor{blue}{(a)}. We choose the convention that the $\nu=1$ integer channel propagates in the positive $x$ direction, while the $\nu=1/3$ fractional channel propagates in the negative $x$ direction. To describe charge equilibration along the line junction, we use the discrete network depicted in Fig.~\ref{fig:setup}\textcolor{blue}{(a)}. We divide the line junction into $N$ cells of length $a$, such that $L=Na$, and label the cells by an integer $i$ with $1\leq i\leq N$.

\subsection{Voltage profile}\label{subsec:voltageprofile}
We begin with deriving the voltage profile along the line junction. The incoming voltages of the $i$th cell, depicted in the yellow, dashed box in Fig.~\ref{fig:setup}\textcolor{blue}{(a)}, are denoted $V_{1,i}$ for the integer channel and $V_{3,i+1}$ for the fractional channel. The the corresponding outgoing voltages are $V_{1,i+1}$ and $V_{3,i}$. Current conservation in the $i$th tunneling link gives the constraint
\begin{align} \label{eq:currentconserve}
    I_{1, i+1} = I_{1, i}  - I_{\tau,i}\,, \quad I_{3, i+1} = I_{3, i} - I_{\tau,i} \,,
\end{align}
where $I_{1,i}$ and $I_{3,i}$ are the currents along the $i$th segment of 
the $\nu = 1$ integer and $\nu = 1/3$ fractional channels, respectively. They are associated with the local voltages $V_{1,i}$
and $V_{3, i}$ as 
\begin{align}
    I_{1,i} = G_0 V_{1,i}\,, \quad  I_{3,i} = \frac{G_0}{3} V_{3,i}\,. 
\end{align}
The tunneling current $I_{\tau,i}$ from the integer channel to the fractional channel is parameterized by a dimensionless conductance $g_{\text{ch}}$ as 
\begin{align} \label{eq:tunnelingcurrent}
    I_{\tau,i} = g_{\text{ch}} G_0(V_{1,i} - V_{3,i+1})\,,
\end{align}
where $g_{\text{ch}}$ is, for simplicity, assumed to be uniform in space and independent of energy over the relevant voltage window, see Ref.~\cite{Nosiglia2018} for a detailed discussion of this assumption. 

We next take the continuum limit $a \rightarrow 0$, keeping the charge-equilibration length $\ell_{\rm{ch}} \equiv a / g_{\text{ch}}$ fixed. By combining Eqs.~\eqref{eq:currentconserve}-\eqref{eq:tunnelingcurrent}, we obtain the relations
\begin{subequations}
\begin{align}
    V_{1,i+1} - V_{1,i} &= - g_{\text{ch}} \left(V_{1,i}-V_{3,i+1}\right), \\ V_{3,i+1}-V_{3,i}
&=
-3g_{\text{ch}} \left(V_{1,i}-V_{3,i+1}\right).
\end{align}
\end{subequations}
Since $V_{3,i+1}=V_{3,i}+O(a)$, the voltage difference in the tunneling current can be evaluated at the same position in the continuum limit. Writing $x=ia$, this limit gives
\begin{align} \label{eq:voltagediffeq}
    \frac{d}{dx} \begin{pmatrix}
        V_1 \\ V_{3} 
    \end{pmatrix} = \frac{1}{\ell_{\rm{ch}}} \begin{pmatrix}
        - 1 & 1 \\ -3 & 3
    \end{pmatrix}\begin{pmatrix}
        V_1 \\ V_{3} 
    \end{pmatrix}\,, 
\end{align}
where we have defined the characteristic charge equilibration length $\ell_{\text{ch}} \equiv a/g_{\text{ch}}$. 

\begin{figure}[t]
 \centering
 \includegraphics[width=0.9\columnwidth]{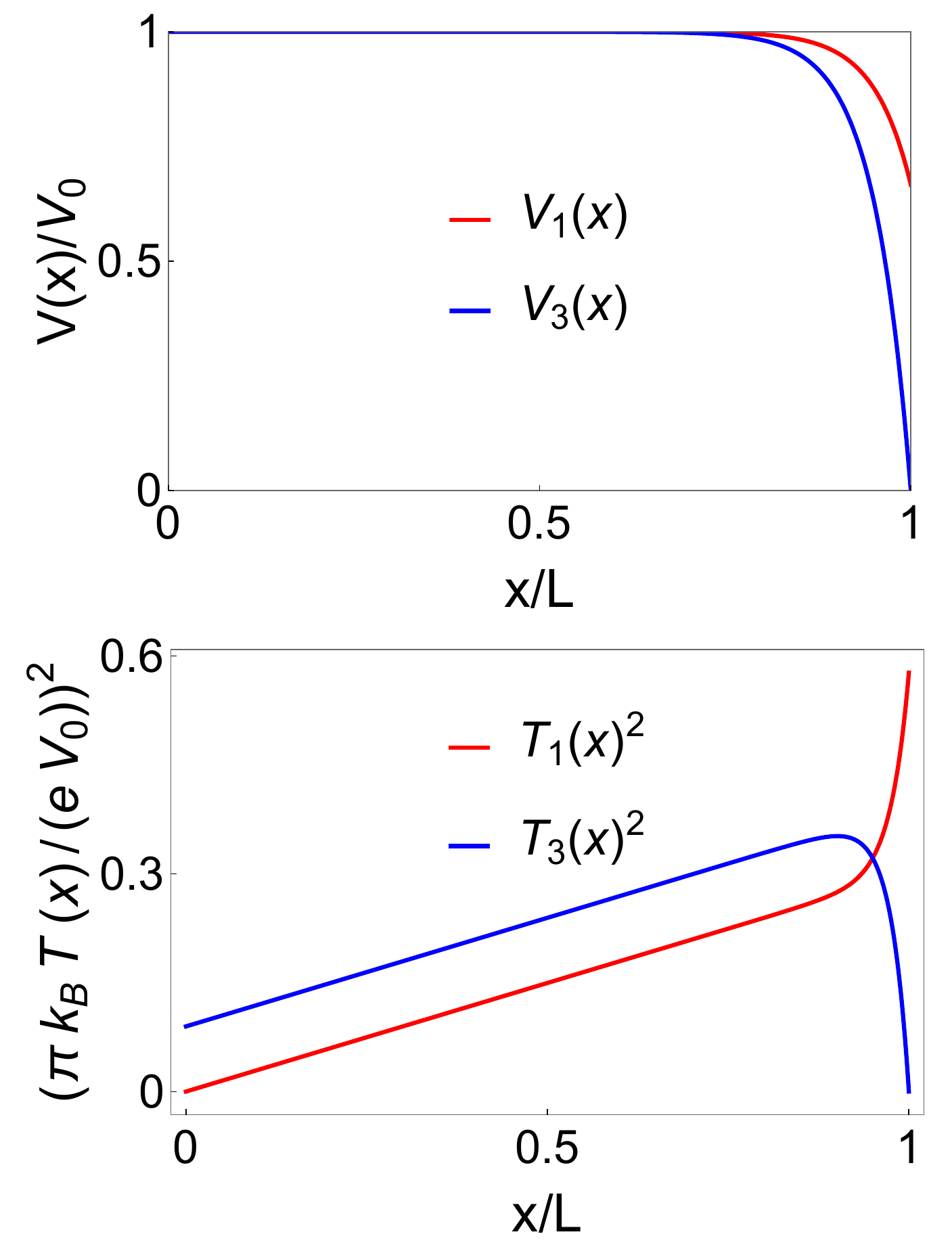}
 \caption{
Voltage and temperature profiles of an equilibrating $\nu=2/3$ line junction for $V_L=V_0$, $V_R=0$, and $T_0=0$. 
(a) Local voltage profiles $V_1(x)$ and $V_3(x)$, normalized by $V_0$. 
The voltage difference between the two channels is strongly suppressed over most of the line junction and remains appreciable only near $x=L$, where the upstream fractional channel is injected. 
(b) Corresponding temperature profiles, normalized as $(\pi k_B T_{1,3}(x)/eV_0)^2$. 
The Joule heating generated near $x=L$ is transported toward the opposite end, $x=0$, through thermal equilibration between the counter-propagating channels, resulting in approximately linear profiles of $T_{1,3}^2$ away from the heating region. For the plot, we use $L/\ell_{\rm ch}=10$ and $\ell_{\rm th}/\ell_{\rm ch}=3$.
}
 \label{fig:voltagetemperatureprofile}
 \end{figure}

We solve the differential equation~\eqref{eq:voltagediffeq} with the boundary conditions
$V_1(0)=V_L$ and $V_3(L)=V_R$, which thus specifies the voltages of the two incoming modes. The resulting voltage profiles read
\begin{subequations}
\label{eq:voltage_profile_linejunction}
\begin{align}
V_1(x)
&=
\frac{
3e^{2L/\ell_{\rm ch}}V_L-V_R
-
e^{2x/\ell_{\rm ch}}(V_L-V_R)
}{
3e^{2L/\ell_{\rm ch}}-1
}\,,\\
V_3(x)
&=
\frac{
3e^{2L/\ell_{\rm ch}}V_L-V_R
-
3e^{2x/\ell_{\rm ch}}(V_L-V_R)
}{
3e^{2L/\ell_{\rm ch}}-1
}\,.
\end{align}
\end{subequations}
We plot these two voltage profiles in Fig.~\ref{fig:voltagetemperatureprofile}\textcolor{blue}{(a)} for $V_L=V_0$ and $V_R=0$. 
In the strong charge-equilibration regime, $\ell_{\rm ch}\ll L$, the voltage profiles are constant at $V_0$ along most of the line junction. In this region, the local voltage difference $V_1(x)-V_3(x)$ is exponentially suppressed, indicating that the two counter-propagating channels are almost fully charge equilibrated. An appreciable voltage difference develops only within a distance of order $\ell_{\rm ch}$ from $x=L$, where is where the upstream fractional channel is injected. Thus, both the voltage drop and the inter-channel voltage imbalance mainly occur near this end of the line junction.

Although the local voltages vary along the line junction, the total
current  in the positive $x$ direction,
\begin{align}
\label{eq:I_tot}
I(x)
\equiv I_1(x) - I_{3}(x)=
G_0
\Big[
V_1(x)-\frac{1}{3}V_3(x)
\Big]\,,
\end{align}
is independent of $x$. Indeed, substituting the profiles~\eqref{eq:voltage_profile_linejunction} into Eq.~\eqref{eq:I_tot}, we find
\begin{align}
I
=
G_1(L)V_L
-
G_3(L)V_R\,,
\label{eq:linejunction_current}
\end{align}
where 
\begin{subequations}
\label{eq:linejunction_conductances}
\begin{align}
G_1(L)
&=
\frac{2}
{3-e^{-2L/\ell_{\rm ch}}}G_0\,,
\\
G_3(L)
&=
\frac{2e^{-2L/\ell_{\rm ch}}}
{3\left(3-e^{-2L/\ell_{\rm ch}}\right)}G_0\,.
\end{align}
\end{subequations}
The conductances $G_1(L)$ and $G_3(L)$
describe the line junction response to the injected voltages $V_L$ and $V_R$, respectively. 

In the strong charge-equilibration regime, $\ell_{\rm ch}\ll L$, the conductances~\eqref{eq:linejunction_conductances}
reduce to
\begin{align}
G_1(L)
&\simeq
\frac{2}{3}G_0\,,
\qquad
G_3(L)
\simeq
0\,,
\label{eq:linejunction_conductances_long}
\end{align}
up to small exponential corrections $\sim e^{-2L/\ell_{\text{ch}}}$. The strong asymmetry between $G_1(L)$ and $G_3(L)$ originates from the net chirality of the edge. For voltage injection through the downstream integer channel, $G_1(L)$ approaches the equilibrated conductance $2G_0/3$, whereas the response to voltage injection through the upstream fractional channel becomes exponentially suppressed, $G_3(L)\simeq0$.

To make contact with the conventional two-terminal conductance, we next consider a Hall-bar geometry in which a second $\nu=2/3$ edge with the opposite chirality connects the same left and right reservoirs. The two edges then contribute oppositely propagating incoming modes, and the total current driven by the voltage difference $V_L-V_R$ is
\begin{align}
I_{\rm 2T}
=
G(L)(V_L-V_R)\,,
\end{align}
where the two-terminal conductance is
\begin{align}
G(L)
&=
G_1(L)+G_3(L)
=
\frac{2}{3}
\frac{3+e^{-2L/\ell_{\rm ch}}}
{3-e^{-2L/\ell_{\rm ch}}}
G_0\,.
\label{eq:two_terminal_conductance}
\end{align}
In the strong charge-equilibration limit, $\ell_{\rm ch}\ll L$, the two-terminal conductance approaches the quantized value $G(L)\simeq 2G_0/3$.

\subsection{Temperature profile}
\label{subsec:temperatureprofile}
We now consider the temperature profile along the line junction. We define the local temperatures $T_{1,i}$ and $T_{3,i}$ for the integer and fractional channels, respectively. The $i$th cell connects the incoming temperatures $T_{1,i}$ and $T_{3,i+1}$ to the outgoing temperatures $T_{1,i+1}$ and $T_{3,i}$, respectively. Energy conservation in this cell gives
\begin{align}
\label{eq:local_J_currents}
J_{1,i+1}
&=
J_{1,i}
-
J_{\tau,i},
\quad
J_{3,i+1}
=
J_{3,i}
-
J_{\tau,i},
\end{align}
where $J_{1,i}$ and $J_{3,i}$ denote the energy currents carried by the two channels in their respective propagation directions. Each local energy current has two contributions: one associated with the electrochemical potential and the other carried by thermal excitations. They are given by
\begin{subequations}
\begin{align}
J_{1,i}
&=
\frac{G_0}{2}V_{1,i}^2
+
\frac{\pi^2 k_B^2}{6h}T_{1,i}^2,
\\
J_{3,i}
&=
\frac{G_0}{6}V_{3,i}^2
+
\frac{\pi^2 k_B^2}{6h}T_{3,i}^2.
\end{align}
\end{subequations}
The energy current transferred between the two channels in the $i$th cell contains contributions from both charge transfer and thermal energy exchange. We parametrize it as
\begin{align}
J_{\tau,i}
&=
\frac{g_{\text{ch}} G_0}{2}
(
V_{1,i}^2
-
V_{3,i+1}^2
)
+
\frac{g_{\rm th}\pi^2 k_B^2}{6h}
(
T_{1,i}^2
-
T_{3,i+1}^2
),
\label{eq:energy_tunneling}
\end{align}
where $g_{\rm th}$ describes thermal equilibration between the two channels, while $g_{\text{ch}}$ parametrizes the charge equilibration defined in Eq.~\eqref{eq:tunnelingcurrent}. The first term in Eq.~\eqref{eq:energy_tunneling} follows from the electrical energy carried by the tunneling particles. For an energy-independent tunneling conductance, particles within a voltage interval $dV$ contribute a charge current $(g_{\rm ch}G_0)dV$ and carry an energy $eV$ per particle. The corresponding energy current is therefore obtained by integrating $(g_{\rm ch}G_0)VdV$ from $V_{3,i+1}$ to $V_{1,i}$, yielding the first term of Eq.~\eqref{eq:energy_tunneling}. Correspondingly, we define the thermal equilibration length as
$
\ell_{\rm th}
\equiv
a/ g_{\rm th},
$
in analogy with $\ell_{\rm ch}=a/g_{\text{ch}}$.

We now take the continuum limit $a\rightarrow0$, keeping $\ell_{\rm ch} \equiv a/g_{\text{ch}}$ and $\ell_{\rm th} \equiv a/g_{\text{th}}$ fixed. 
Using the voltage equations derived in Sec.~\ref{subsec:voltageprofile}, together with the energy conservation in Eq.~\eqref{eq:local_J_currents} and Eq.~\eqref{eq:energy_tunneling}, we obtain
\begin{align}
\frac{d (T_1^2)}{dx}
&=
\frac{3e^2}{\pi^2 k_B^2\ell_{\rm ch}}
\left(V_1-V_3\right)^2
-
\frac{T_1^2-T_3^2}{\ell_{\rm th}},
\nonumber \\
\frac{d (T_3^2)}{dx}
&=
-
\frac{3e^2}{\pi^2 k_B^2\ell_{\rm ch}}
\left(V_1-V_3\right)^2
-
\frac{T_1^2-T_3^2}{\ell_{\rm th}}.
\label{eq:temperature_diffeq}
\end{align}
Here, the terms $\propto(V_1-V_3)^2$ describe the Joule heating generated by the charge equilibration,  whereas the terms proportional to $T_1^2-T_3^2$ describe thermal equilibration between the two counter-propagating channels.

We now solve Eq.~\eqref{eq:temperature_diffeq} using the voltage profile in Eqs.~\eqref{eq:voltage_profile_linejunction} and the boundary conditions
$T_1(0)=T_L = T_0$ and $T_3(L)=T_R = T_0$. The resulting temperature profiles read 
\begin{align}
T_1^2(x)
&=
T_0^2
+
\Theta_V
\bigg[
\Big(
\frac{1}{2}
-
\frac{\ell_{\rm ch}}{4\ell_{\rm th}}
\Big)
\big(
e^{4x/\ell_{\rm ch}}-1
\big)
\nonumber \\
&
+
\frac{x}{L+\ell_{\rm th}}
\Big \{
1+
\left(
\frac{1}{2}
+
\frac{\ell_{\rm ch}}{4\ell_{\rm th}}
\right)
\big(
e^{4L/\ell_{\rm ch}}-1
\big)
\Big\}
\bigg]\,,
\nonumber \\
T_3^2(x)
&=
T_0^2
+
\Theta_V
\bigg[
\Big(
\frac{1}{2}
+
\frac{\ell_{\rm ch}}{4\ell_{\rm th}}
\Big)
\big(
e^{4L/\ell_{\rm ch}}
-
e^{4x/\ell_{\rm ch}}
\big)
\nonumber \\
&
-
\frac{L-x}{L+\ell_{\rm th}}
\Big\{
1+
\Big(
\frac{1}{2}
+
\frac{\ell_{\rm ch}}{4\ell_{\rm th}}
\Big)
\big(
e^{4L/\ell_{\rm ch}}-1
\big)
\Big\}
\bigg]\,,
\label{eq:temperature_profiles}
\end{align}
where 
\begin{align}
\Theta_V
\equiv
\frac{6e^2(V_L-V_R)^2}
{\pi^2 k_B^2
\left(3e^{2L/\ell_{\rm ch}}-1\right)^2}\,
\label{eq:thetaV}
\end{align}
parametrizes the total Joule-heating occurring in the line junction.

The temperature profiles~\eqref{eq:temperature_profiles} is plotted in Fig.~\ref{fig:voltagetemperatureprofile}\textcolor{blue}{(b)}.
In the strong charge-equilibration regime, $\ell_{\rm ch}\ll L$, the voltage drop and the inter-channel voltage imbalance mainly occur near $x=L$. The Joule heating is therefore localized near this end of the junction. The generated heat can then flow toward the opposite end, $x=0$, through energy exchange between the counter-propagating channels. Away from the narrow Joule-heating region, both $T_1^2$ and $T_3^2$ vary approximately linearly with position, which is a characteristic of diffusive heat transport along the equilibrating edge. The extent of this inter-channel heat exchange is controlled by the ratio $\ell_{\rm th}/L$. Importantly, heat can therefore reach the region near $x=0$ even though the local voltage difference there is strongly suppressed. This feature will play a central role for the noise in the line junction, which we now turn to.

\subsection{Current fluctuations and noise}
Here, we model the the current fluctuations and noise generated in the single line junction. We focus on sufficiently low frequencies such that charge accumulation within each cell can be neglected. In this low-frequency limit, not only the average current but also its fluctuations obey current conservation in each equilibration cell. For the $i$th cell, current conservation leads to 
\begin{align}
\delta I_{1,i+1}(t)
&=
\delta I_{1,i}(t) -
\delta I_{\tau,i}(t) \,. 
\nonumber \\
\delta I_{3,i+1}(t)
&=
\delta I_{3,i}(t) -
\delta I_{\tau,i}(t)\,.
\label{eq:fluctuation_current_conservation}
\end{align}
The fluctuations of the currents carried by the two bare channels can further be expressed in terms of the corresponding local voltage fluctuations as
\begin{align}
\delta I_{1,i}
&=
G_0\delta V_{1,i}\,,
\qquad
\delta I_{3,i}
=
\frac{G_0}{3}\delta V_{3,i}\,.
\label{eq:current_voltage_fluctuation}
\end{align}
Moreover, the fluctuation of the tunneling current in each cell can be decomposed into two contributions. In the $i$th cell, the fluctuation of the tunneling current reads 
\begin{align}
\delta I_{\tau,i} (t)
= G_0 g_{\rm ch}
\left(
\delta V_{1,i}(t)
-
\delta V_{3,i+1}(t)
\right)
+
\xi_i (t)\,
\label{eq:tunneling_current_fluctuation}
\end{align}
The first term in Eq.~\eqref{eq:tunneling_current_fluctuation} has its origin in the fluctuations of the local voltage difference across the equilibration link. By contrast, the second term represents the intrinsic current fluctuations $\xi_i(t)$ associated with the tunneling process itself~\cite{Park_Noise_2019}.
The intrinsic fluctuations in different cells are taken to be statistically independent and are characterized by a local noise kernel,
\begin{align}
\left\langle
\xi_i(\omega)
\xi_j(\omega')
\right\rangle
\equiv
2\pi \delta(\omega+\omega')
\delta_{ij}
{\mathcal S}_{\tau,i}(\omega)\,.
\label{eq:intrinsic_noise_kernel}
\end{align}
Here, $\xi_i(\omega)$ is the Fourier transform of $\xi_i(t)$. At this stage, we leave the local noise kernel ${\mathcal S}_{\tau,i}(\omega)$ unspecified. Its microscopic form depends on the nature of the equilibration process and, in general, on the local voltage and temperature differences between the two channels. A specific form of the local noise kernel will be introduced when evaluating the resulting expressions below.

Using Eqs.~\eqref{eq:fluctuation_current_conservation}--\eqref{eq:tunneling_current_fluctuation}, defining the intrinsic tunneling-current fluctuation density $\xi(x,t)\equiv\xi_{i}(t)/a$, and taking the continuum limit, one can derive the following differential equations for voltage-fluctuation profiles in the line junction 
\begin{align}
    \frac{d}{dx} \begin{pmatrix}
        \delta V_{1} \\ \delta V_{3} 
    \end{pmatrix} = \frac{1}{\ell_{\text{ch}}} \begin{pmatrix}
        - 1 & 1 \\ -3 & 3
    \end{pmatrix}\begin{pmatrix}
        \delta V_{1}  \\ \delta V_{3}
    \end{pmatrix} - \frac{\xi(x, t)}{G_0} \begin{pmatrix}
    1 \\ 3
    \end{pmatrix}\,,
    \label{eq:voltage_fluctuation_diffeq}
\end{align}
We now solve Eq.~\eqref{eq:voltage_fluctuation_diffeq} by specifying the fluctuations of the two incoming modes, $\delta V_1(0,\omega)$ and $\delta V_3(L,\omega)$. 
The voltage fluctuation of the outgoing fractional channel at $x=0$ is then given by
\begin{align}
\delta V_3(0,\omega)
&= \frac{1}{3e^{2L/\ell_{\rm ch}}-1} \bigg (
3(e^{2L/\ell_{\rm ch}}-1)\delta V_1(0,\omega)
\nonumber \\ &+
2\delta V_3(L,\omega)+\frac{6}{G_0}
\int_0^L dx\,e^{2(L-x)/\ell_{\rm ch}}\xi(x,\omega) \bigg )\,, 
\label{eq:V3_fluctuation_x0}
\end{align}
Thus, the outgoing voltage fluctuation consists of the fluctuations carried by the two incoming modes and the intrinsic fluctuations generated throughout the line junction.

Since the two channels are counter-propagating, the total current flowing in the positive $x$ direction is $I=I_1-I_3$. The current fluctuation at $x=0$ thus reads
\begin{align} 
\delta I(\omega)
&=
G_0\delta V_1(0,\omega)
-
\frac{G_0}{3}\delta V_3(0,\omega)
\nonumber \\
&=
\frac{2}{3e^{2L/\ell_{\rm ch}}-1}
\Big[
G_0
\Big(
e^{2L/\ell_{\rm ch}} \delta V_1(0,\omega)-
\frac{1}{3}\delta V_3(L,\omega)
\Big) \nonumber \\ 
& \qquad -
\int_0^L dx e^{2(L-x)/\ell_{\rm ch}}\xi(x,\omega)
\Big].
\label{eq:linejunction_current_fluctuation}
\end{align}
Using the expression for the noise kernel \eqref{eq:intrinsic_noise_kernel}, we obtain the resulting current noise $S_{I}(\omega) 
\equiv \langle \delta I(\omega) \delta I(-\omega)\rangle$, 
\begin{align} 
S_I(\omega)
&=
\frac{4}{(3e^{2L/\ell_{\rm ch}}-1)^2}
\Big[
G_0^2
\Big(
e^{4L/\ell_{\rm ch}} S_{V_1}^{\rm in}(\omega)
+
\frac{1}{9}S_{V_3}^{\rm in}(\omega)
\Big) \nonumber \\ 
& \qquad +
\int_0^L dx e^{4(L-x)/\ell_{\rm ch}}
{\mathcal S}_{\tau}(x,\omega)
\Big],
\label{eq:linejunction_noise}
\end{align}
where $S_{{V}_m}^{\rm in}(\omega)\equiv
\langle\delta V_m(\omega)\delta V_m(-\omega)\rangle$ with $m=1,3$, and ${\mathcal S}_{\tau}(x,\omega)\equiv{\mathcal S}_{\tau,i}(\omega)/a$ is the noise density generated by the tunneling links of the line junction. At sufficiently low frequencies, charge accumulation along the line junction can be neglected, and the current fluctuations are conserved. Consequently, $S_I(\omega)$ is independent of the position at which the current is evaluated.

We now apply Eq.~\eqref{eq:linejunction_noise} to some particular cases. We first consider the equilibrium limit, where the two channels have the same ambient temperature $T_0$ and no voltage difference. The low-frequency voltage fluctuations of the incoming modes are then given by
\begin{align}
S_{V_1}^{\rm in}(0)
&=
\frac{2k_B T}{G_0}\,,
\qquad
S_{V_3}^{\rm in}(0)
=
\frac{6k_B T}{G_0}\,,
\label{eq:JNnoise_incomingvoltage}
\end{align}
while the local, low-frequency tunneling-noise density 
$S_{\tau}(x, \omega)$  is given by 
\begin{align}
    \mathcal{S}_\tau (x, \omega = 0) = \frac{4G_0 k_B T_0}{
    \ell_{\text{ch}}
    } \,. \label{eq:tunnelingnoisedensityJN}
\end{align}
Substituting these expressions \eqref{eq:JNnoise_incomingvoltage} and \eqref{eq:tunnelingnoisedensityJN} into Eq.~\eqref{eq:linejunction_noise}, we obtain
\begin{align}
S_I^{\rm eq}(0)
=
\frac{4 G_0 k_B T_0}{3}
\frac{3e^{2L/\ell_{\rm ch}}+1}{3e^{2L/\ell_{\rm ch}}-1} = 2 k_B T_0 G(L)\,,
\label{eq:linejunction_equilibrium_noise}
\end{align}
reproducing the Johnson--Nyquist relation. Here $G (L)$ is the two-terminal conductance \eqref{eq:two_terminal_conductance}. 

We next investigate how voltage fluctuations injected from the boundary propagate though the line junction. We thus set $\xi(x,\omega)=0$ and Eq.~\eqref{eq:linejunction_current_fluctuation} then gives
\begin{align}
\delta I(\omega)
=
G_1(L)\delta V_1(0,\omega)
-
G_3(L)\delta V_3(L,\omega)\,,
\label{eq:boundary_voltage_to_current}
\end{align}
where the conductances $G_1(L)$ and $G_3(L)$ are given in Eqs.~\eqref{eq:linejunction_conductances}. 
If the voltage fluctuations injected from the two boundaries are statistically independent, the corresponding current noise is
\begin{align}
S_I(\omega)
=
G_1^2(L)S_{V_1}^{\rm in}(\omega)
+
G_3^2(L)S_{V_3}^{\rm in}(\omega)\,.
\label{eq:boundary_voltage_noise}
\end{align}
In the strong charge-equilibration regime, $\ell_{\rm ch}\ll L$, the propagation of the boundary voltage fluctuations becomes strongly asymmetric. A voltage fluctuation injected in the downstream integer channel at $x=0$ is transmitted as current fluctuations with the equilibrated conductance $G_1(L)\simeq 2G_0/3$. In contrast, a voltage fluctuation injected through the upstream fractional channel at $x=L$ gives an exponentially small contribution to the current noise since $G_3(L)\simeq 0$. Thus, fluctuations injected from the downstream side of the line junction strongly dominate over those injected from the upstream side, reflecting the net chirality of the edge. 

Finally, we consider the noise generated by a finite dc voltage difference $V_L-V_R$. 
To focus on the bias-induced noise, we set the base temperature to zero, $T_0=0$, such that the voltage fluctuations injected from the two reservoirs vanish. 
The current noise in Eq.~\eqref{eq:linejunction_noise} is then entirely determined by the intrinsic fluctuations generated in the equilibration links along the line junction.

An important feature of Eq.~\eqref{eq:linejunction_noise} is its strong spatial asymmetry, as discussed previously in Refs.~\cite{Park_Noise_2019, Spanslatt_Noise_2019}. 
In the strong charge-equilibration regime, $\ell_{\rm ch}\ll L$, the weighting factor in Eq.~\eqref{eq:linejunction_noise} is exponentially concentrated near $x=0$, where the downstream integer channel is injected. 
Fluctuations generated in this region therefore contribute dominantly to the measured current noise. 
The location of this dominant noise-generating region is fixed by the net chirality of the edge and is ultimately governed by the bulk topological order of the FQH state. 
We refer to such regions as ``noise-generating spots'' throughout this paper.

Interestingly, as shown in Fig.~\ref{fig:voltagetemperatureprofile}\textcolor{blue}{(a)}, the local voltage difference $V_1-V_3$ is strongly suppressed at the noise-generating spot, whereas the Joule heating produced by the applied dc bias is concentrated near the opposite end, $x=L$, where the voltage difference remains appreciable. 
As discussed in Sec.~\ref{subsec:temperatureprofile}, part of this heat is transported toward $x=0$ through energy exchange between the counter-propagating channels. 
The local temperatures at the noise-generating spot can therefore remain finite even though the local voltage difference there is exponentially suppressed. 
Consequently, in the low-frequency regime, the local noise kernel near the noise-generating spot is dominated by its Johnson--Nyquist contribution,
\begin{align}
{\mathcal S}_{\tau}(x,\omega)
\simeq
\frac{2G_0 k_B}{\ell_{\rm ch}}
\left[
T_1(x)+T_3(x)
\right]\,.
\label{eq:localJohnsonNyquist}
\end{align}
Thus, a finite dc bias generates excess noise through a two-step process: Joule heating is produced near $x=L$, and this heat is transported to the noise-generating spot near $x=0$, where it produces thermal current fluctuations. This mechanism will be particularly important when we apply the line-junction result to the four-terminal QPC geometry below.

In the diffusive thermal regime, $\ell_{\rm ch}\ll \ell_{\rm th}\ll L$, the heat transported from the Joule-heating region gives rise to a finite temperature at the noise-generating spot. As a result, Eqs.~\eqref{eq:linejunction_noise} and \eqref{eq:localJohnsonNyquist} yield the characteristic excess-noise scaling for the combination of ballistic charge flow and diffusive heat flow,
\begin{align}
S_I(0)
\sim
G_0 e|V_L-V_R|
\sqrt{\frac{\ell_{\rm th}}{L}}\,,
\label{eq:linejunction_noise_scaling}
\end{align}
This result reproduces the characteristic scaling obtained for an equilibrating $\nu=2/3$ line junction in Refs.~\cite{Park_Noise_2019, Spanslatt_Noise_2019}. Here, this result will serve as the building block for the four-terminal geometry considered below, where line-junction noise generated in different arms contributes differently to the auto- and cross-current correlations measured in the drain contacts.

\section{Four-terminal QPC geometry}
\label{sec:fourterminalQPC}
Having established the equilibration properties of a single line junction, we now move on to the four-terminal geometry shown in Fig.~\ref{fig:setup}\textcolor{blue}{(b)}. 
This geometry consists of four equilibrating $\nu=2/3$ line junctions, denoted by UL, UR, DL, and DR, which are connected through a single QPC located at $x=0$. 
Each arm has length $L$. 
We choose a coordinate $x$, with $-L\leq x\leq L$, increasing from left to right, such that the left contacts UL and DL are located at $x=-L$, while the right contacts UR and DR are located at $x=L$. 
With this convention, the downstream integer channel propagates in the positive $x$ direction along the upper edge and in the negative $x$ direction along the lower edge. The fractional channel always propagate in the opposite direction as the integer channel in the same arm.

We consider the situation where a dc voltage $V_0$ is applied to the UL contact, whereas the DR contact is grounded. The current fluctuations are measured at the drain contacts UR and DL. In the following, we connect the line-junction solutions derived in Sec.~\ref{sec:linejunction} by imposing the matching conditions at the QPC and the boundary conditions at the four contacts.

\subsection{Voltage and temperature profiles}
\label{subsec:voltagetempprofiles_QPC}
We first determine the voltage profiles in the four arms. Using the coordinate $x$ introduced above, the four line-junction equations for the voltages, on the form~\eqref{eq:voltagediffeq}, can be written as
\begin{align}
\frac{d}{dx}
\begin{pmatrix}
V_{1,A} \\
V_{3,A}
\end{pmatrix}
=
\frac{\sigma_A}{\ell_{\rm ch}}
\begin{pmatrix}
-1 & 1 \\
-3 & 3
\end{pmatrix}
\begin{pmatrix}
V_{1,A} \\
V_{3,A}
\end{pmatrix}\,,
\label{eq:voltagediffequpperlower}
\end{align}
where $A={\rm UL},{\rm UR},{\rm DL},{\rm DR}$ and
$\sigma_{\rm UL}=\sigma_{\rm UR}=-\sigma_{\rm DL}=-\sigma_{\rm DR}=1$. The boundary conditions corresponding to the chosen bias configuration are
\begin{align}
V_{1,{\rm UL}}(-L)
&=
V_0\,,
\quad
V_{1,{\rm DR}}(L)
=
0\,.
\label{eq:source_boundary_voltage}
\end{align}
The drain contacts UR and DL are treated as floating contacts. The current entering each floating contact is fully re-emitted into the outgoing FQH edge channels, such that no net current is absorbed by the contact. Assuming full equilibration at the contact, the outgoing channels are characterized by a common electrochemical potential. This leads to the boundary conditions
\begin{align}
V_{1,{\rm UR}}(L)
&=
V_{3,{\rm UR}}(L)\,,
\quad
V_{1,{\rm DL}}(-L)
=
V_{3,{\rm DL}}(-L)\,.
\label{eq:boundaryconditionvoltage}
\end{align}

The local temperature equations for the four arms are obtained analogously from Eq.~\eqref{eq:temperature_diffeq}. Using $\sigma_A$ introduced in Eq.~\eqref{eq:voltagediffequpperlower}, we write
\begin{align}
\frac{d}{dx}
\begin{pmatrix}
T_{1,A}^2 \\[1mm]
T_{3,A}^2
\end{pmatrix}
&=
\sigma_A \frac{3e^2}{\pi^2 k_B^2\ell_{\rm ch}}
\left(V_{1,A}-V_{3,A}\right)^2
\begin{pmatrix}
1 \\ -1
\end{pmatrix}
\nonumber \\
&\qquad
-
\frac{\sigma_A}{\ell_{\rm th}}
\begin{pmatrix}
1 & -1 \\
1 & -1
\end{pmatrix}
\begin{pmatrix}
T_{1,A}^2 \\[1mm]
T_{3,A}^2
\end{pmatrix}\,.
\label{eq:temperature_diffeq_fourarms}
\end{align}
All contacts are assumed to act as ideal thermal reservoirs at the base temperature $T_0$. The temperatures of the modes injected from the contacts therefore satisfy
\begin{align}
T_{1,{\rm UL}}(-L)
=
T_{3,{\rm UR}}(L)
=
T_{3,{\rm DL}}(-L)
=
T_{1,{\rm DR}}(L)
=
T_0\,.
\label{eq:temperature_contact_bc}
\end{align}

The four arms are connected at the QPC located at $x=0$, see Fig.~\ref{fig:setup}\textcolor{blue}{(b)}. 
We assume that the integer channel fully passes through the QPC without inter-edge tunneling, whereas the $\nu=1/3$ channel is partially transmitted with inter-edge transmission probability $\mathcal{T}$. 
The voltages at the QPC are then related by
\begin{subequations}
\label{eq:voltagesQPC}
\begin{align}
V_{1,{\rm UR}}(0)
&=
V_{1,{\rm UL}}(0),
\quad
V_{1,{\rm DL}}(0)
=
V_{1,{\rm DR}}(0), \\
V_{3,{\rm UL}}(0)
&=
(1-\mathcal{T})V_{3,{\rm UR}}(0)
+
\mathcal{T}V_{3,{\rm DL}}(0),
\\
V_{3,{\rm DR}}(0)
&=
\mathcal{T}V_{3,{\rm UR}}(0)
+
(1-\mathcal{T})V_{3,{\rm DL}}(0).
\end{align}
\end{subequations}
Similarly, energy conservation at the QPC gives  
\begin{subequations}
\label{eq:QPCboundarycond_temp}
\begin{align}
T_{1,{\rm UR}}^2(0)
&=
T_{1,{\rm UL}}^2(0),
\qquad
T_{1,{\rm DL}}^2(0)
=
T_{1,{\rm DR}}^2(0),\\
T_{3,{\rm UL}}^2(0)
&=
(1-\mathcal{T})T_{3,{\rm UR}}^2(0)
+
\mathcal{T}T_{3,{\rm DL}}^2(0)
\nonumber\\
&+
\frac{e^2 \mathcal{T}(1-\mathcal{T})}{\pi^2 k_B^2}
\left[
V_{3,{\rm UR}}(0)-V_{3,{\rm DL}}(0)
\right]^2, \label{eq:qpc_temperature_fractionalUL} \\
T_{3,{\rm DR}}^2(0)
&=
\mathcal{T} T_{3,{\rm UR}}^2(0)
+
(1-\mathcal{T}) T_{3,{\rm DL}}^2(0)
\nonumber \\
&+
\frac{e^2 \mathcal{T}(1-\mathcal{T})}{\pi^2 k_B^2}
\left[
V_{3,{\rm UR}}(0)-V_{3,{\rm DL}}(0)
\right]^2.
\label{eq:qpc_temperature_fractionalDR}
\end{align}
\end{subequations}
The last terms in Eqs.~\eqref{eq:qpc_temperature_fractionalUL} and \eqref{eq:qpc_temperature_fractionalDR} represent the Joule heating generated by partitioning at the QPC.

Equations~\eqref{eq:voltagediffequpperlower}, together with the contact boundary conditions in Eqs.~\eqref{eq:source_boundary_voltage} and \eqref{eq:boundaryconditionvoltage} and the QPC matching conditions in Eq.~\eqref{eq:voltagesQPC}, determine the voltage profiles in all four arms. 
Once the voltage profiles are obtained, the temperature profiles follow from Eq.~\eqref{eq:temperature_diffeq_fourarms}, supplemented by the thermal boundary conditions in Eq.~\eqref{eq:temperature_contact_bc} and the energy-conservation conditions at the QPC, Eqs.~\eqref{eq:qpc_temperature_fractionalUL} and \eqref{eq:qpc_temperature_fractionalDR}.

An interesting feature of the four-terminal geometry is a chirality-induced back-action effect: Inter-edge tunneling between the fractional channels at the QPC changes the voltage of the outgoing $\nu=1/3$ channel. 
As this $\nu=1/3$ channel propagates further along the edge, it equilibrates with the integer channel, and in the steady state, a part of the current is fed back towards the QPC. This feedback loop effectively modifies the local voltage difference incident on the QPC itself. 
As a result, the tunneling current is determined self-consistently rather than simply by the bare transmission probability $\mathcal{T}$. In the strong charge-equilibration limit, $L\gg\ell_{\rm ch}$, the voltage difference across the QPC, $\Delta V_{\text{QPC}}$, and the local voltage discontinuity of the $\nu = 1/3$ channel at the QPC become
\begin{subequations}
\begin{align}
&\Delta V_{\text{QPC}} \equiv V_{3,{\rm UR}}(0)-V_{3,{\rm DL}}(0)
\simeq
\frac{V_0}{1+\mathcal{T}}, \label{eq:voltagedifferenceQPC}
\\
&V_{3,{\rm UR}}(0)-V_{3,{\rm UL}}(0)
=
V_{3,{\rm DR}}(0)-V_{3,{\rm DL}}(0) \nonumber 
\\ & \qquad \qquad \qquad \qquad \,\,\,
\simeq
\frac{\mathcal{T}}{1+\mathcal{T}}V_0.
\label{eq:voltagedropQPC}
\end{align}
\end{subequations}
Accordingly, the tunneling charge current at the QPC reads
\begin{align}
I_{\mathcal T}
&=
\frac{G_0}{3}\mathcal{T}
\left[
V_{3,{\rm UR}}(0)-V_{3,{\rm DL}}(0)
\right]
\simeq
\frac{G_0V_0}{3}
\frac{\mathcal{T}}{1+\mathcal{T}}.
\label{eq:tunnelingcurrent_QPC}
\end{align}
Thus, the chirality-induced back action effectively renormalizes the QPC transmission from $\mathcal{T}$ to
\begin{align} \label{eq:effectivetransmission}
\mathcal{T}_{\rm eff}
\equiv
\frac{\mathcal{T}}{1+\mathcal{T}}.
\end{align}

\begin{figure}[t]
 \centering
\includegraphics[width=1.0\columnwidth]{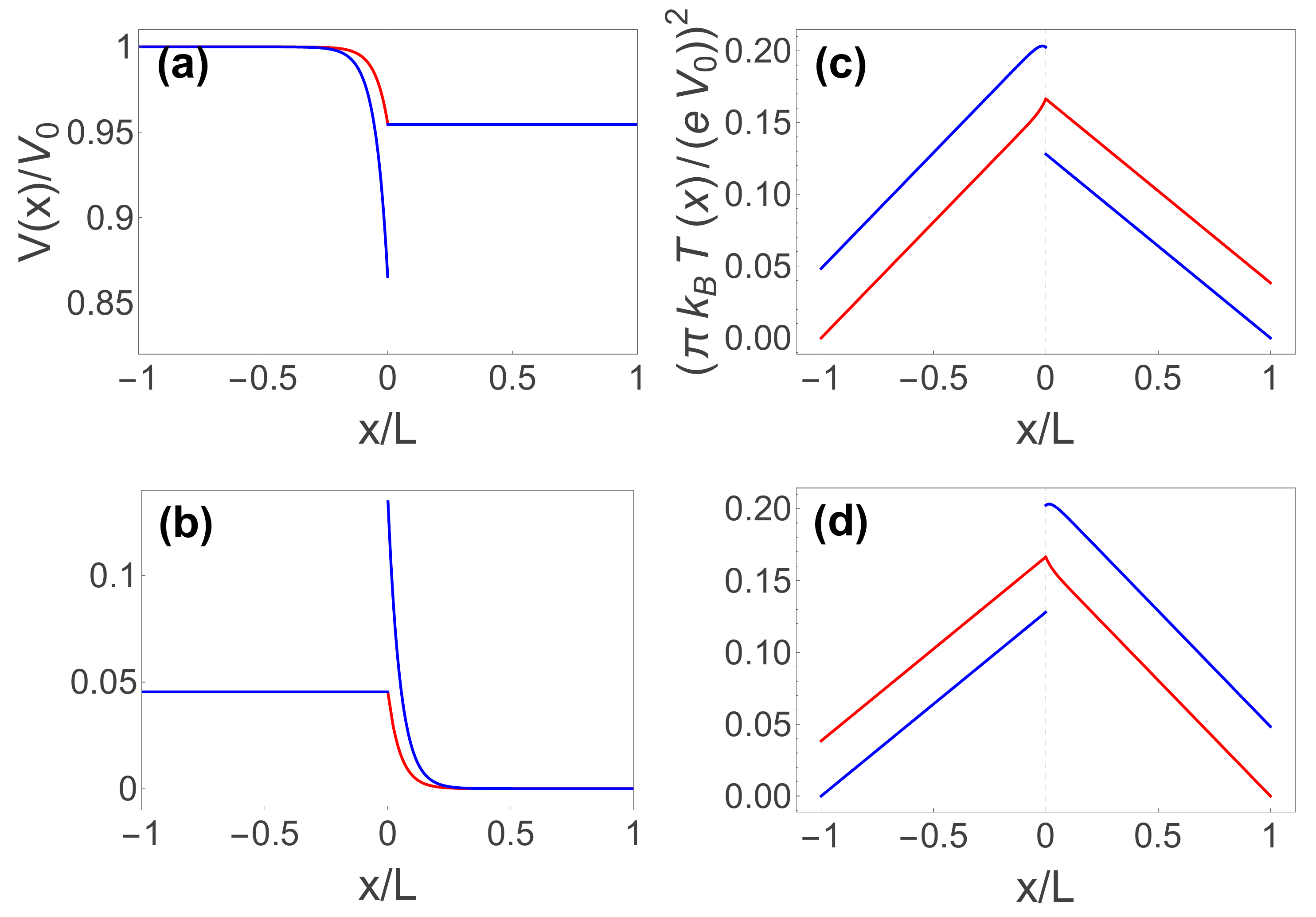}
\caption{Voltage and temperature profiles along the upper and lower edges of the four-terminal QPC geometry. We use $L/\ell_{\rm ch}=10$, $\ell_{\rm th}/\ell_{\rm ch}=3$, $T_0=0$, and $\mathcal{T}=0.1$.
(a) Voltage profiles along the upper edge, from the UL contact at $x=-L$ to the UR contact at $x=L$.
(b) Corresponding voltage profiles along the lower edge, from the DL contact to the DR contact.
(c) Temperature profiles along the upper edge.
(d) Corresponding temperature profiles along the lower edge.
The red and blue curves represent the integer and fractional channels, respectively.
The voltages are normalized by $V_0$, while the temperatures are shown as $(\pi k_B T/eV_0)^2$.
The vertical dashed line at $x=0$ indicates the position of the QPC. 
}

 \label{fig:voltage_temperature_profile_QPC}
 \end{figure}

The same inter-edge partitioning generates also Joule heating at the QPC. The corresponding total heat power generated at the QPC is
\begin{align}
P_{\text{QPC}}
&=
\frac{G_0}{3}\mathcal{T}(1-\mathcal{T})
\left[
V_{3,{\rm UR}}(0)-V_{3,{\rm DL}}(0)
\right]^2
\nonumber\\
&\simeq
\frac{G_0V_0^2}{3}
\frac{\mathcal{T}(1-\mathcal{T})}
{(1+\mathcal{T})^2},
\label{eq:heatpowerQPC}
\end{align}
where in the second line, we used the voltage difference in Eq.~\eqref{eq:voltagedifferenceQPC} in the strong charge-equilibration limit. 
The generated heat is equally distributed between the two outgoing fractional channels, each carrying $P_{\text{QPC}}/2$.

The resulting voltage profiles along the upper and lower edges are shown in Figs.~\ref{fig:voltage_temperature_profile_QPC}\textcolor{blue}{(a)-(b)}. 
In the strong charge-equilibration regime, the voltage drop and the inter-channel voltage difference occurs within a distance of order $\ell_{\rm ch}$ on the upstream side of the QPC along both edges (i.e., $-\ell_{\rm ch}\lesssim x<0$ along the upper edge and $0<x\lesssim\ell_{\rm ch}$ along the lower edge). 
Away from these regions, the integer and fractional channels are nearly fully charge equilibrated to the same voltages. At the QPC, the integer-channel voltage remains continuous, whereas the fractional-channel voltage exhibits the discontinuity given in Eq.~\eqref{eq:voltagedropQPC} which arises from tunneling at the QPC. Note that no additional voltage drop occurs near the drain contacts due to the floating boundary conditions imposed at the drains.

The corresponding temperature profiles are shown in Figs.~\ref{fig:voltage_temperature_profile_QPC}\textcolor{blue}{(c)-(d)}. The temperature profiles are symmetric between the upper and lower edges. This follows from the symmetric heat generation at the QPC together with the identical base temperature $T_0$ of all the boundary contacts. In particular, the two incoming fractional channels at the QPC have the same temperature,
$T_{3,{\rm UR}}(0)=T_{3,{\rm DL}}(0)$.
Consequently, the power dissipated at the QPC, $P_{\rm QPC}$ in Eq.~\eqref{eq:heatpowerQPC}, gives rise directly to a discontinuity in $T_3^2$ across the QPC. Since each outgoing fractional channel carries $P_{\rm QPC}/2$, the QPC boundary condition~\eqref{eq:QPCboundarycond_temp} gives
\begin{align}
&T_{3,{\rm UL}}^2(0)-T_{3,{\rm UR}}^2(0)
=
T_{3,{\rm DR}}^2(0)-T_{3,{\rm DL}}^2(0)
\nonumber\\
&=
\frac{3h}{\pi^2 k_B^2}P_{\rm QPC}
\simeq
\frac{e^2V_0^2}{\pi^2 k_B^2}
\frac{\mathcal{T}(1-\mathcal{T})}{(1+\mathcal{T})^2}.
\end{align}
The resulting jump in $T_3^2$ is clearly visible in Fig.~\ref{fig:voltage_temperature_profile_QPC}\textcolor{blue}{(c)-(d)}.

\subsection{Current fluctuations and noise correlations}
\label{subsec:autocrosscorrelation}

We now turn to the current fluctuations and the corresponding auto- and cross-current correlations measured at the drain contacts UR and DL. 
The voltage and temperature profiles obtained in Sec.~\ref{subsec:voltagetempprofiles_QPC} determine the local noise sources along the four arms, while partitioning at the QPC provides an additional noise source. 
We first determine how these fluctuations propagate through the four-terminal geometry and then evaluate their contributions to the auto- and cross-correlations at the drains. We then analyze the resulting expressions in the weak-tunneling regime, $\mathcal T\ll1$, and show that the cross-correlation noise is more insensitive to the noise generated by heating along the arms than the auto-correlation. We finally construct a Fano factor $F_{\rm clean}$ that further suppresses the heating contribution and allows a more reliable extraction of the quasiparticle charge from tunneling between complex edges.

\begin{figure}[t]
 \centering
\includegraphics[width=1.0\columnwidth]{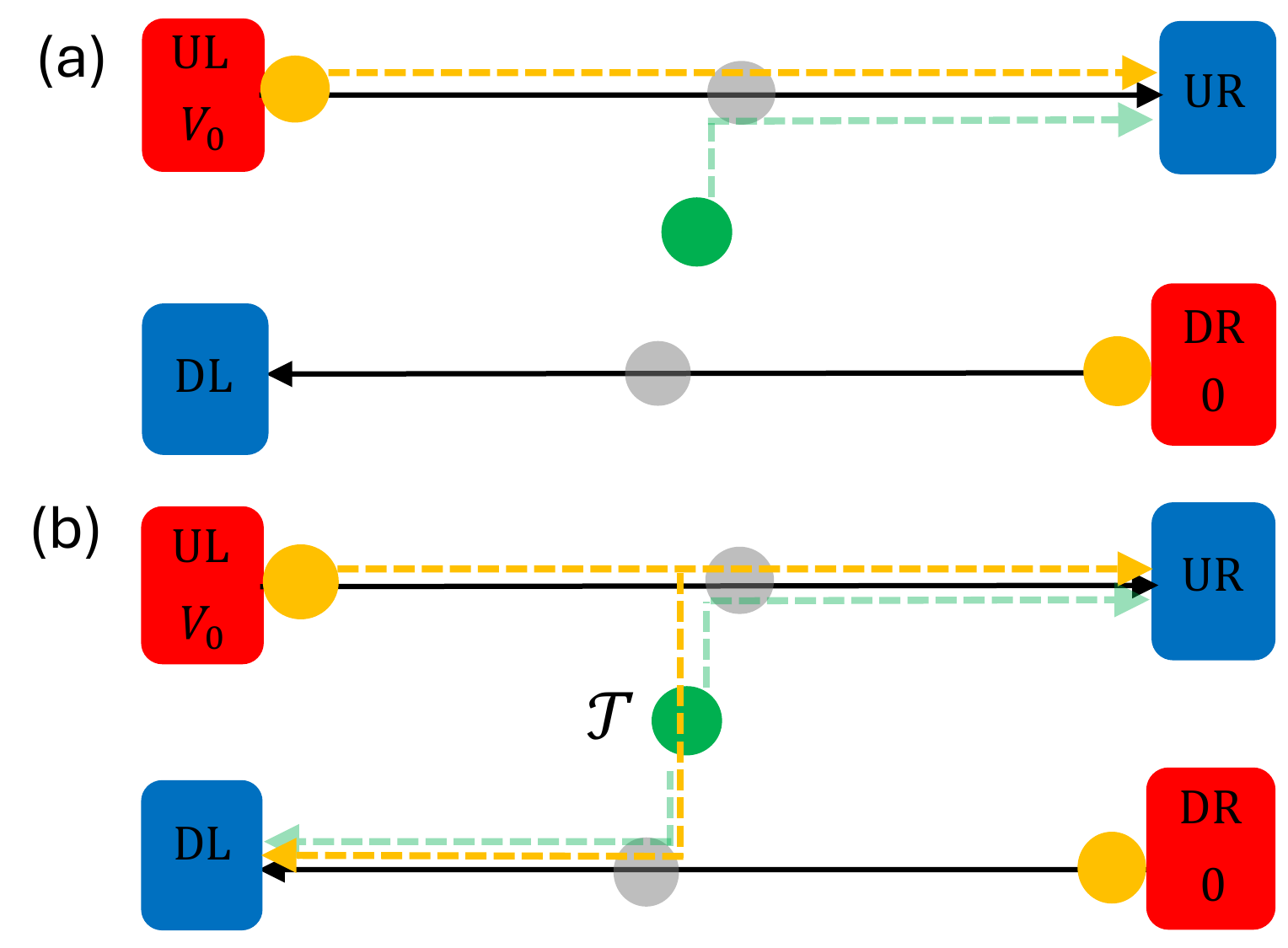}
\caption{Schematic illustrations of how noise generated at different locations
contributes to the drain-current correlations. The black solid lines represent the edges, with the arrows indicating
the dominating, downstream propagation directions.
The yellow, gray, and green spots represent the outer-arm, inner-arm,
and QPC noise-generating spots, respectively, while the colored
dashed arrows indicate the propagation of the corresponding current
fluctuations.
(a) For the auto-correlation, noise generated in an outer arm can
propagate directly to the corresponding drain.
(b) For the cross-correlation, the same outer-arm fluctuation contributes
to the opposite drain only after transmission through the QPC, indicated
by $\mathcal T$, thereby suppressing the arm-heating contribution.
}
 \label{fig:Noisegeneratingspots}
 \end{figure}

We begin by considering current fluctuations generated in the arms of the QPC setup.
Following the analysis of a single equilibrating line junction in Sec.~\ref{sec:linejunction}, we introduce an intrinsic tunneling-current fluctuation density $\xi_A(x,t)$ in each arm $A={\rm UL},{\rm UR},{\rm DL},{\rm DR}$.
We then apply the single-line-junction result in Eq.~\eqref{eq:linejunction_current_fluctuation} to each of the four arms.
Adapting it to our coordinate convention of $-L \leq x \leq L$, we obtain
\begin{subequations}
\label{eq:current_fluctuation_fourarms}
\begin{align}
\delta I_{\rm UL}
&=
G_1(L)\delta V_{1,{\rm UL}}(-L)
-
G_3(L)\delta V_{3,{\rm UL}}(0)
+
\delta I_{\rm UL}^{\rm int},
\\
\delta I_{\rm UR}
&=
G_1(L)\delta V_{1,{\rm UR}}(0)
-
G_3(L)\delta V_{3,{\rm UR}}(L)
+
\delta I_{\rm UR}^{\rm int},
\\
\delta I_{\rm DL}
&=
G_1(L)\delta V_{1,{\rm DL}}(0)
-
G_3(L)\delta V_{3,{\rm DL}}(-L)
+
\delta I_{\rm DL}^{\rm int},
\\
\delta I_{\rm DR}
&=
G_1(L)\delta V_{1,{\rm DR}}(L)
-
G_3(L)\delta V_{3,{\rm DR}}(0)
+
\delta I_{\rm DR}^{\rm int}.
\end{align}
\end{subequations}
Here, $\delta I_{A}^{\text{int}}$ denote the intrinsic contributions generated along the four arms, written as
\begin{align}
\delta I_A^{\rm int}
=
-
\frac{G_1(L)}{G_0}
\int_A dx\,
e^{-2d_A(x)/\ell_{\rm ch}}\xi_A(x),
\label{eq:intrinsic_fluctuation_fourarms}
\end{align}
where $d_A(x)$ denotes the distance from the point where the integer channel enters arm $A$,
\begin{align}
d_{\rm UL}(x)&=x+L,\qquad
d_{\rm UR}(x)=x,\nonumber\\
d_{\rm DL}(x)&=-x,\qquad
d_{\rm DR}(x)=L-x.
\end{align}
Equation~\eqref{eq:intrinsic_fluctuation_fourarms} identifies the regions
from which intrinsic fluctuations contribute most strongly to the current
in each arm, i.e., the ``noise-generating spots.'' Specifically, these
spots are located near $x=-L$ for UL, $x=0$ for UR and DL, and $x=L$
for DR, as indicated in
Fig.~\ref{fig:Noisegeneratingspots}. We refer to the noise-generating
spots near the QPC in the UR and DL arms as the inner-arm spots (gray spots Fig.~\ref{fig:Noisegeneratingspots}), and
those near the outer contacts in the UL and DR arms as the outer-arm
spots (yellow spots).

To determine how these fluctuations contribute to the currents measured at the two drains, we next impose the fluctuation boundary conditions at the boundary contacts and at the QPC. 
At a finite base temperature $T_0$, we retain the equilibrium voltage fluctuations injected from the UL and DR contacts. 
We denote these independent boundary fluctuations by
\begin{align}
\delta V_{\rm UL}^{\rm in}
&\equiv
\delta V_{1,{\rm UL}}(-L),
&
\delta V_{\rm DR}^{\rm in}
&\equiv
\delta V_{1,{\rm DR}}(L).
\label{eq:boundaryconditionQPC0}
\end{align}
In contrast, the voltage fluctuations at the floating UR and DL contacts are determined self-consistently by the floating boundary conditions,
\begin{align}
\delta V_{1,{\rm UR}}(L)
&=
\delta V_{3,{\rm UR}}(L),
\quad
\delta V_{1,{\rm DL}}(-L)
=
\delta V_{3,{\rm DL}}(-L).
\label{eq:boundaryconditionQPC1}
\end{align}
At the QPC, the integer-channel voltage fluctuations remain continuous,
\begin{align}
\delta V_{1,{\rm UR}}(0)
&=
\delta V_{1,{\rm UL}}(0),
&
\delta V_{1,{\rm DL}}(0)
&=
\delta V_{1,{\rm DR}}(0),
\label{eq:boundaryconditionQPC2}
\end{align}
whereas the fractional-channel fluctuations satisfy
\begin{align}
\delta V_{3,{\rm UL}}(0)
&=
(1-\mathcal T)\delta V_{3,{\rm UR}}(0)
+
\mathcal T\delta V_{3,{\rm DL}}(0)
+
\frac{3}{G_0}\xi_{\mathcal T},
\nonumber\\
\delta V_{3,{\rm DR}}(0)
&=
\mathcal T\delta V_{3,{\rm UR}}(0)
+
(1-\mathcal T)\delta V_{3,{\rm DL}}(0)
-
\frac{3}{G_0}\xi_{\mathcal T},
\label{eq:qpc_fluctuation_bc}
\end{align}
where $\xi_{\mathcal T}$ denotes the intrinsic current fluctuations generated at the QPC.

We now assume strong charge-equilibration, $\ell_{\rm ch}\ll L$, and solve Eq.~\eqref{eq:current_fluctuation_fourarms} with the boundary conditions~\eqref{eq:boundaryconditionQPC0}--\eqref{eq:qpc_fluctuation_bc}. We then obtain the drain fluctuations as
\begin{subequations}
\label{eq:drain_current_fluctuations}
\begin{align}
\delta I_{\rm UR}
&=
\frac{1}{2(1+\mathcal T)}
\Big[
(2+\mathcal T)\delta I_{\rm UL}^{\rm int}
+
3\mathcal T\delta I_{\rm UR}^{\rm int}
-
3\mathcal T\delta I_{\rm DL}^{\rm int}
\nonumber\\
&
+
\mathcal T\delta I_{\rm DR}^{\rm int}
+
2\xi_{\mathcal T}
+
\frac{2G_0}{3}
\left[
(2+\mathcal T)\delta V_{\rm UL}^{\rm in}
+
\mathcal T\delta V_{\rm DR}^{\rm in}
\right]
\Big],
\\
\delta I_{\rm DL}
&=
\frac{1}{2(1+\mathcal T)}
\Big[(2+\mathcal T)\delta I_{\rm DR}^{\rm int}
-
3\mathcal T\delta I_{\rm UR}^{\rm int}
+
3\mathcal T\delta I_{\rm DL}^{\rm int}
\nonumber\\
&+\mathcal T\delta I_{\rm UL}^{\rm int}
-
2\xi_{\mathcal T}
+
\frac{2G_0}{3}
\left[
\mathcal T\delta V_{\rm UL}^{\rm in}
+
(2+\mathcal T)\delta V_{\rm DR}^{\rm in}
\right]
\Big].
\end{align}
\end{subequations}

To evaluate the auto- and cross-correlations of the drain-current fluctuations, $\delta I_{\text{UR}}$ and $\delta I_{\text{DL}}$, we define the intrinsic noise generated in arm $A$ and at the QPC as
\begin{align}
S_A^{\rm int}
&\equiv
\left\langle
\delta I_A^{\rm int}(\omega)
\delta I_A^{\rm int}(-\omega)
\right\rangle,
\\
S_{\mathcal T}
&\equiv
\left\langle
\xi_{\mathcal T}(\omega)
\xi_{\mathcal T}(-\omega)
\right\rangle .
\end{align}
The equilibrium voltage fluctuations injected from the UL and DR contacts satisfy
\begin{align}
S_V^{\rm eq}
&\equiv
\left\langle
\delta V_{\rm UL}^{\rm in}(\omega)
\delta V_{\rm UL}^{\rm in}(-\omega)
\right\rangle
=
\left\langle
\delta V_{\rm DR}^{\rm in}(\omega)
\delta V_{\rm DR}^{\rm in}(-\omega)
\right\rangle
\nonumber\\
&=
\frac{2k_BT_0}{G_0}.
\label{eq:contact_equilibrium_noise}
\end{align}
The intrinsic noise sources in different arms, the QPC noise, and the equilibrium fluctuations injected from the contacts are statistically independent, so that the cross-correlations between distinct noise sources vanish.

In the strong charge-equilibration regime, $L\gg \ell_{\rm ch}$, the
temperature profiles are symmetric under the transformations
$\mathrm{UL}\leftrightarrow\mathrm{DR}$ and
$\mathrm{UR}\leftrightarrow\mathrm{DL}$, as shown in
Figs.~\ref{fig:voltage_temperature_profile_QPC}\textcolor{blue}{(c)-(d)}.
Accordingly, the intrinsic noises generated in the corresponding arms satisfy
\begin{align}
S_{\rm UL}^{\rm int}
&=
S_{\rm DR}^{\rm int}
\equiv S_{\rm out}^{\rm int},
&
S_{\rm UR}^{\rm int}
&=
S_{\rm DL}^{\rm int}
\equiv S_{\rm in}^{\rm int}.
\label{eq:arm_noise_symmetry}
\end{align}
Here, $S_{\rm out}^{\rm int}$ and $S_{\rm in}^{\rm int}$ denote the
intrinsic noises associated with the outer-arm and inner-arm
noise-generating spots, respectively.
The two auto-correlations are therefore identical, and we obtain
\begin{align}
S_{\rm auto}
&=
S_{\rm UR,UR}
=
S_{\rm DL,DL} \equiv \langle \delta I_{\text{DL}} (\omega) \delta I_{\text{DL}} (-\omega) \rangle
\nonumber\\
&=
\frac{(2+\mathcal T)^2+\mathcal T^2}
{4(1+\mathcal T)^2}
S_{\rm out}^{\rm int}
+
\frac{9\mathcal T^2}
{2(1+\mathcal T)^2}
S_{\rm in}^{\rm int}
+
\frac{S_{\mathcal T}}
{(1+\mathcal T)^2}
\nonumber\\
&\quad+
\frac{G_0^2}{9}
\frac{(2+\mathcal T)^2+\mathcal T^2}
{(1+\mathcal T)^2}
S_V^{\rm eq}.
\label{eq:noise_auto}
\end{align}
Similarly, the cross-correlation between the two drain currents is
\begin{align}
S_{\rm cross}
&\equiv
S_{\rm UR,DL}
=
S_{\rm DL,UR}\equiv \langle \delta I_{\text{DL}} (\omega) \delta I_{\text{UR}} (-\omega) \rangle
\nonumber\\
&=
\frac{\mathcal T(2+\mathcal T)}
{2(1+\mathcal T)^2}
S_{\rm out}^{\rm int}
-
\frac{9\mathcal T^2}
{2(1+\mathcal T)^2}
S_{\rm in}^{\rm int}
-
\frac{S_{\mathcal T}}
{(1+\mathcal T)^2}
\nonumber\\
&\quad+
\frac{2G_0^2}{9}
\frac{\mathcal T(2+\mathcal T)}
{(1+\mathcal T)^2}
S_V^{\rm eq}.
\label{eq:noise_cross}
\end{align}
The auto-correlation and cross-correlation noises have contributions three different origins: the intrinsic noise generated by equilibration in the arms, denoted $S_{\text{out}}^{\text{int}}$ and $S_{\text{in}}^{\text{int}}$, the intrinsic noise generated by partitioning of quasiparticles in the QPC, $S_{\mathcal{T}}$, and the equilibrium noise injected by the contacts, $S_V^{\text{eq}}$. 

The different coefficients of the noise terms in
Eqs.~\eqref{eq:noise_auto} and \eqref{eq:noise_cross} can be
understood from the propagation of the fluctuations illustrated in
Fig.~\ref{fig:Noisegeneratingspots}. Consider first the outer-arm
noises, illustrated by the yellow spots. A fluctuation generated in an
outer arm can propagate directly to the corresponding drain and
therefore contributes to the auto-correlation with a coefficient of
order unity. In contrast, its contribution to the opposite drain
requires transmission through the QPC, as illustrated in
Fig.~\ref{fig:Noisegeneratingspots}\textcolor{blue}{(b)}. Consequently, the outer-arm
contribution to the cross-correlation acquires an additional factor of
$\mathcal T$.

The inner-arm noise, indicated by the gray spots, behaves differently.
For a fluctuation generated near the QPC in the UR or DL arm to
contribute to the zero-frequency drain-current noise, the QPC must be
partially open. Otherwise, the particle-hole excitations generated at
an inner-arm noise spot propagate downstream to the same drain, so
that their contributions cancel in the dc current fluctuation.
Finite transmission through the QPC separates these excitations between
the two drains, giving a fluctuation amplitude proportional to
$\mathcal T$. Consequently, the corresponding contribution to both the
auto- and cross-correlations is proportional to $\mathcal T^2$.
The coefficients of this contribution enter the two drain currents with
opposite signs, resulting in a negative contribution to the
cross-correlation. 

Finally, the QPC noise source, indicated by the green spot, directly
generates partitioning fluctuations that propagate toward the two
drains with opposite signs. It thus contributes positively to
the auto-correlation and negatively to the cross-correlation, as seen
in Eqs.~\eqref{eq:noise_auto} and \eqref{eq:noise_cross}.

We next express the intrinsic noise sources entering
Eqs.~\eqref{eq:noise_auto} and \eqref{eq:noise_cross} in terms of the
local voltages and temperatures. For the QPC, we assume a finite-temperature
partition-noise form
\begin{align}
S_{\mathcal T}
&=
\frac{2G_0}{3}\,
\mathcal T(1-\mathcal T)\,
e^*|\Delta V_{\rm QPC}|
\coth\left(
\frac{e^*|\Delta V_{\rm QPC}|}{2k_BT_{\rm in}}
\right),
\label{eq:qpc_noise}
\end{align}
for a quasiparticle charge $e^*$ transferred across the QPC. In Eq.~\eqref{eq:qpc_noise}, $\Delta V_{\text{QPC}}$ denotes the local voltage difference across the QPC given in Eq.~\eqref{eq:voltagedifferenceQPC}, and $T_{\rm in}$ denotes the local temperature of the
two fractional channels incident on the QPC, 
\begin{align}
T_{\rm in}
&\equiv
T_{3,\rm UR}(0)
=
T_{3,\rm DL}(0)\,. 
\label{eq:Tin_definition}
\end{align}
In the absence of equilibration along the edge, the noise generated in the arms is absent, while $\Delta V_{\rm QPC}\to V_0$ and $T_{\rm in}\to T_0$. Equation~\eqref{eq:qpc_noise} then reduces to the well-known, finite-temperature QPC partition-noise expression~\cite{Feldman2017Mar}.
By contrast, in the
strong charge-equilibration regime, and to leading order in $\mathcal T$,
the local temperature becomes
\begin{align}
T_{\rm in}^2
=
T_0^2+
\frac{e^2V_0^2}{\pi^2k_B^2}
\frac{L^2}
{\ell_{\rm th}(2L+\ell_{\rm th})}
\mathcal T
+O(\mathcal T^2).
\label{eq:Tin_strong}
\end{align}

Equation~\eqref{eq:qpc_noise} is exact for noninteracting fermions upon
replacing $e^*\mapsto e$, whereas for FQH quasiparticles we employ it as a
phenomenological interpolation. It reproduces the expected equilibrium
limit, $\Delta V_{\rm QPC}\rightarrow 0$, and the shot-noise limit,
$T_{\rm in}\rightarrow 0$, but does not incorporate the interaction-dependent crossover associated with the scaling dimension
of the quasiparticle tunneling operator in the chiral Luttinger-liquid
description of FQH edges~\cite{Snizhko2015,Veillon2024,Ruelle2025}.
Nevertheless, this form has been successfully used to describe noise
measurements in FQH experiments for Laughlin quasiparticles~\cite{dePicciotto1997,Saminadayar_fractional_charge_1997,Heiblum2020edge},
and is sufficient for the purpose of the present analysis.

The intrinsic noise generated along each arm can be obtained directly
from the result for a single equilibrating line junction,
Eq.~\eqref{eq:localJohnsonNyquist}. In the strong
charge-equilibration regime, $L\gg\ell_{\rm ch}$, the spatial kernel
entering the arm noise is localized near the corresponding
noise-generating spot of each arm [cf. Eqs.~\eqref{eq:intrinsic_fluctuation_fourarms}], and the intrinsic noise is thus
determined by the local temperatures of the two counterpropagating
channels. Applying Eq.~\eqref{eq:localJohnsonNyquist} to these
noise-generating spots illustrated in Fig.~\ref{fig:Noisegeneratingspots}, we obtain
\begin{align}
S_{\rm out}^{\rm int}
&=
\frac{2G_0k_B}{9}
\left(T_0+T_{\rm out}\right),
\label{eq:Sout_temperature}
\\
S_{\rm in}^{\rm int}
&=
\frac{2G_0k_B}{9}
\left(T_{1,\rm in}+T_{\rm in}\right).
\label{eq:Sin_temperature}
\end{align}
Here, $T_{\rm out}$ is the local temperature of the fractional channel
at the outer-arm spots as depicted as the yellow dots, while
$T_{1,\rm in}$ is the local temperature of the integer channel at the
inner-arm spots as depicted as the gray dots:
\begin{align}
T_{\rm out}
&\equiv
T_{3,{\rm UL}}(-L)
=
T_{3,{\rm DR}}(L),
\\
T_{1,\rm in}
&\equiv
T_{1,{\rm UR}}(0)
=
T_{1,{\rm DL}}(0).
\end{align}
In the strong charge-equilibration regime, to leading order in
$\mathcal T$, these temperatures are given by
\begin{align}
T_{\rm out}^2
&=
T_0^2+
\frac{e^2V_0^2}{\pi^2k_B^2}
\frac{L+\ell_{\rm th}}
{2L+\ell_{\rm th}}
\mathcal T
+O(\mathcal T^2),
\label{eq:Tout_strong}
\\
T_{1,\rm in}^2
&=
T_0^2+
\frac{e^2V_0^2}{\pi^2k_B^2}
\frac{L(L+\ell_{\rm th})}
{\ell_{\rm th}(2L+\ell_{\rm th})}
\mathcal T
+O(\mathcal T^2).
\label{eq:T1in_strong}
\end{align}
In particular, in the long-arm limit $L\gg\ell_{\rm th}$,
\begin{align}
T_{\rm out}^2
\simeq
T_0^2+
\frac{e^2V_0^2}{2\pi^2k_B^2}\mathcal T
+O(\mathcal T^2),
\label{eq:Tout_long}
\end{align}
so that the heating of the fractional channel at the outer
noise-generating spots remains finite even for arbitrarily long arms.
As we show below, this surviving outer-arm heating gives a substantial
contribution to the auto-correlation noise in the large-bias regime.

Before discussing the nonequilibrium contributions, it is useful to verify
that Eqs.~\eqref{eq:noise_auto} and \eqref{eq:noise_cross} reproduce the
correct equilibrium noise. At equilibrium, $V_0=0$, all channels are at the base temperature $T_0$,
and thus substituting $T_{1,A}(x) = T_{3,A}(x) = T_0$ into the line-junction result \eqref{eq:linejunction_noise} gives 
\begin{align}
S_{\rm out}^{\rm int, \text{eq}}
=
S_{\rm in}^{\rm int, \text{eq}}
=
\frac{4G_0k_BT_0}{9}.
\label{eq:arm_noise_equilibrium}
\end{align}
Furthermore, the QPC noise \eqref{eq:qpc_noise} reduces to
\begin{align}
S_{\mathcal T}^{\rm eq}
=
\frac{4G_0k_BT_0}{3}
\mathcal T(1-\mathcal T).
\label{eq:qpc_noise_equilibrium}
\end{align}
Substituting Eqs.~\eqref{eq:arm_noise_equilibrium},
\eqref{eq:qpc_noise_equilibrium}, and
\eqref{eq:contact_equilibrium_noise} into
Eqs.~\eqref{eq:noise_auto} and \eqref{eq:noise_cross}, we obtain
\begin{align}
S_{\rm auto}^{\rm eq}
&=
\frac{4G_0k_BT_0}{3}
=
2k_BT_0 G (L),
\label{eq:auto_equilibrium}
\\
S_{\rm cross}^{\rm eq}
&=0,
\label{eq:cross_equilibrium}
\end{align}
where $G(L)=2G_0/3$ is the two-terminal conductance of the equilibrated $\nu = 2/3$ edges. Thus, the auto-correlation correctly reproduces the Johnson--Nyquist noise, while the equilibrium cross-correlation vanishes.

Having established that the noise expressions reproduce the correct
equilibrium limit, we now focus on the nonequilibrium contributions by
defining the excess auto- and cross-correlation noises as
\begin{align}
\Delta S_{\rm auto}
&\equiv
S_{\rm auto}-S_{\rm auto}^{\rm eq},
\\
\Delta S_{\rm cross}
&\equiv
S_{\rm cross}-S_{\rm cross}^{\rm eq}
=
S_{\rm cross},
\end{align}
and the excess QPC noise as
$
\Delta S_{\mathcal T}
\equiv
S_{\mathcal T}-S_{\mathcal T}^{\rm eq}.
$
For a finite $V_0$ in the weak inter-edge transmission regime,
$\mathcal T\ll1$, expanding the coefficients in
Eqs.~\eqref{eq:noise_auto} and \eqref{eq:noise_cross} to leading order
in $\mathcal T$ gives
\begin{align}
\Delta S_{\rm auto}
&\simeq
\frac{2G_0k_B}{9}
(1-\mathcal T)
\left(T_{\rm out}-T_0\right)
+
\Delta S_{\mathcal T},
\label{eq:excess_auto}
\\
\Delta S_{\rm cross}
&\simeq
\frac{2G_0k_B}{9}
\mathcal T
\left(T_{\rm out}-T_0\right)
-
\Delta S_{\mathcal T}.
\label{eq:excess_cross}
\end{align}
The contribution from $S_{\rm in}^{\rm int}$ is multiplied by a
prefactor of order $\mathcal T^2$ in both correlations and thus
does not contribute to the leading terms. These expressions explicitly
show that both excess noises vanish at equilibrium, when
$T_{\rm out}=T_0$ and $\Delta S_{\mathcal T}=0$.

Equations~\eqref{eq:excess_auto} and \eqref{eq:excess_cross} are
central results of this work and reveal a qualitative difference
between the two correlations. The auto-correlation contains a
contribution proportional to $T_{\rm out}-T_0$, which originates from
the elevated temperature at the outer-arm noise-generating spots. As shown in Eq.~\eqref{eq:Tout_long},
this contribution remains finite in the long-arm limit
$L\gg\ell_{\rm th}$. In the large-bias regime,
$\big(eV_0/(k_BT_0))^2\mathcal T\gg1$, 
the outer-arm temperature behaves as
$
T_{\rm out}-T_0
\simeq
\frac{e|V_0|}{\sqrt{2}\pi k_B}\sqrt{\mathcal T},
$
and hence the heating contribution to the auto-correlation scales as
$\sqrt{\mathcal T}$. It can thus remain sizable even in the
weak-transmission regime $\mathcal{T} \ll 1$. In contrast, the same outer-arm heating
enters the cross-correlation with an additional factor of
$\mathcal T$, giving a contribution proportional to
$\mathcal T^{3/2}$. The cross-correlation is consequently much less
sensitive to this heating contribution.

To quantify how the different sensitivities to arm heating affect
the extraction of the quasiparticle charge, we define
\begin{align}
F_{\rm auto}
&\equiv
\frac{\Delta S_{\rm auto}}{2e|I_{\mathcal T}|},
\quad
F_{\rm cross}
\equiv
-\frac{\Delta S_{\rm cross}}{2e|I_{\mathcal T}|},
\quad
F_{\rm QPC}
\equiv
\frac{\Delta S_{\mathcal T}}{2e|I_{\mathcal T}|}.
\label{eq:Fano_factors}
\end{align}
Using Eqs.~\eqref{eq:tunnelingcurrent_QPC} and \eqref{eq:qpc_noise}, we find
\begin{align}
F_{\rm QPC}
&=
(1-\mathcal T)
\bigg[
\frac{e^*}{e}
\coth\left(
\frac{e^*|\Delta V_{\rm QPC}|}{2k_BT_{\rm in}}
\right)
-
\frac{2k_BT_0}{e|\Delta V_{\rm QPC}|}
\bigg]
\nonumber\\
&\simeq
\frac{e^*}{e}
\coth\bigg[
\frac{\pi}{2}\frac{e^*}{e}
\sqrt{
\frac{\ell_{\rm th}(2L+\ell_{\rm th})}
{L^2\mathcal T}}
\bigg],
\label{eq:FQPC_largebias}
\end{align}
where the second line applies in the large-bias and weak-transmission
regime. Thus, $F_{\rm QPC}\simeq e^*/e$ requires
$\mathcal T\ll\mathcal T_c$, where
\begin{align}
\mathcal T_c
\equiv
\frac{\pi^2}{4}
\left(\frac{e^*}{e}\right)^2
\frac{\ell_{\rm th}(2L+\ell_{\rm th})}{L^2}.
\label{eq:Tcritical}
\end{align}
Therefore, even at large bias, the quasiparticle charge can be directly
extracted from the intrinsic QPC noise only when the transmission is
sufficiently weak. For example, for $e^*=e/3$ and
$L=3\ell_{\rm th}$, $\mathcal T_c\simeq0.21$.
In the following, we consider the regime
$\mathcal T\ll\mathcal T_c$, for which $F_{\rm QPC}\simeq e^*/e$.

In the regime $\mathcal T\ll\mathcal T_c$, where
$F_{\rm QPC}\simeq e^*/e$, we now examine how the arm heating
affects the Fano factors extracted from the measured correlations.
Using Eqs.~\eqref{eq:Tout_strong}, \eqref{eq:excess_auto}, and
\eqref{eq:excess_cross}, in the large-bias regime we obtain
\begin{align}
F_{\rm auto}
&\simeq
\frac{e^*}{e}
+
\frac{1}{3\pi}
\sqrt{
\frac{L+\ell_{\rm th}}
{(2L+\ell_{\rm th})\mathcal T}
},
\label{eq:Fauto_asymptotic}
\\
F_{\rm cross}
&\simeq
\frac{e^*}{e}
-
\frac{1}{3\pi}
\sqrt{
\frac{L+\ell_{\rm th}}
{2L+\ell_{\rm th}}
\mathcal T
}.
\label{eq:Fcross_asymptotic}
\end{align}
The Fano factor extracted from the auto-correlation acquires a heating
correction proportional to $\mathcal T^{-1/2}$ and therefore becomes
diverging as $\mathcal{T}$ is reduced.
In contrast, the heating correction to $F_{\rm cross}$ scales as
$\mathcal T^{1/2}$ and vanishes in the limit of $\mathcal{T} \ll 1$.
Thus, the cross-correlation provides a substantially more robust probe
of the quasiparticle charge than the auto-correlation, as
also suggested in Ref.~\cite{De_HOM_2025}.

Nevertheless, $F_{\rm cross}$ still contains a finite arm-heating
correction at any nonzero $\mathcal T$. For example, for $e^*=e/3$
and $L=3\ell_{\rm th}$, we obtain
$F_{\rm cross}\simeq0.308$ and $0.297$ for
$\mathcal T=0.1$ and $0.2$, respectively, instead of the ideal value
$e^*/e=1/3$. Since in an experiment the transmission cannot be made
arbitrarily small,
it is desirable to construct a Fano factor that eliminates this leading
heating contribution already at finite transmission. We show below that
this can be achieved by an appropriate combination of the auto- and
cross-correlation noises. 

To isolate the heating contribution, we note that adding the
auto- and cross-correlations in Eqs.~\eqref{eq:noise_auto} and
\eqref{eq:noise_cross} exactly cancels the QPC and inner-arm
contributions. The excess noises \eqref{eq:excess_auto} and \eqref{eq:excess_cross} then satisfy the relation, 
\begin{align}
\Delta S_{\rm auto}+\Delta S_{\rm cross}
&=
\Delta S_{\rm out}^{\rm int}
=
\frac{2G_0k_B}{9}
\left(T_{\rm out}-T_0\right),
\label{eq:arm_noise_isolation}
\end{align}
where
$\Delta S_{\rm out}^{\rm int}
\equiv S_{\rm out}^{\rm int}-S_{\rm out}^{\rm int, \text{eq}}$
is the excess noise generated in the outer-arm spots, while $S_{\rm out}^{\rm int, \text{eq}}$ denotes the corresponding equilibrium value given in 
Eq.~\eqref{eq:arm_noise_equilibrium}. Thus, the combination
$\Delta S_{\rm auto}+\Delta S_{\rm cross}$ directly isolates the
heating contribution.

Using Eq.~\eqref{eq:arm_noise_isolation}, we can remove the outer-arm
heating contribution from the cross-correlation and define a new Fano
factor,
\begin{align}
F_{\rm clean}
\equiv
\frac{
-\Delta S_{\rm cross}
+
\mathcal T_{\rm eff}(2-\mathcal T_{\rm eff})
\left(
\Delta S_{\rm auto}+\Delta S_{\rm cross}
\right)/2
}
{2e|I_{\mathcal T}|(1-\mathcal T_{\rm eff})^2}.
\label{eq:Fclean}
\end{align}
Here, $\mathcal T_{\rm eff}\equiv
3 I_{\mathcal T}/(G_0 V_0)$ is the effective QPC transmission
defined in Eq.~\eqref{eq:effectivetransmission}.
Importantly, $F_{\rm clean}$ is constructed entirely from experimentally
accessible quantities, $\Delta S_{\rm cross}$, $\Delta S_{\rm auto}$,
$\mathcal T_{\rm eff}$, and $I_{\mathcal T}$.
Using Eqs.~\eqref{eq:noise_auto} and \eqref{eq:noise_cross}, we obtain
\begin{align}
F_{\rm clean}
=
F_{\rm QPC}
+
\frac{9\mathcal T_{\rm eff}^2}
{4e|I_{\mathcal T}|(1-\mathcal T_{\rm eff})^2}
\Delta S_{\rm in}^{\rm int},
\label{eq:Fclean_result}
\end{align}
where
$\Delta S_{\rm in}^{\rm int}
\equiv S_{\rm in}^{\rm int}-S_{\rm in}^{\rm int, eq}$
is the excess noise generated in the inner-arm spots, while the corresponding equilibrium
value $S_{\rm in}^{\rm int, eq}$ is given in Eq.~\eqref{eq:arm_noise_equilibrium}.
Thus, the outer-arm heating contribution is eliminated exactly from
$F_{\rm clean}$. In the large-bias and weak-transmission regime,
$\Delta S_{\rm in}^{\rm int}\propto\mathcal T^{1/2}$ and
$|I_{\mathcal T}|\propto\mathcal T$, so that
\begin{align}
F_{\rm clean}
=
F_{\rm QPC}
+O(\mathcal T^{3/2}),
\label{eq:Fclean_scaling}
\end{align}
compared with the $O(\mathcal T^{1/2})$ heating correction to
$F_{\rm cross}$. For $\mathcal T\ll\mathcal T_c$, this further gives
$F_{\rm clean}\simeq e^*/e$.

\begin{table}[t]
\caption{
Comparison of the Fano factors in the large-bias and weak-transmission
regime. Here,
$A=(3\pi)^{-1}
\sqrt{(L+\ell_{\rm th})/(2L+\ell_{\rm th})}$.
The last column shows the leading scaling of the arm-heating correction
relative to the intrinsic QPC Fano factor $F_{\rm QPC}$.
}
\label{tab:fano_comparison}
\begin{ruledtabular}
\begin{tabular}{c c c}
Fano factor
& Leading form
& $F-F_{\rm QPC}$ \\[2pt]
\hline
$F_{\rm auto}$
& $F_{\rm QPC}+A\mathcal T^{-1/2}$
& $O(\mathcal T^{-1/2})$
\\
$F_{\rm cross}$
& $F_{\rm QPC}-A\mathcal T^{1/2}$
& $O(\mathcal T^{1/2})$
\\
$F_{\rm clean}$
& $F_{\rm QPC}+O(\mathcal T^{3/2})$
& $O(\mathcal T^{3/2})$
\end{tabular}
\end{ruledtabular}
\end{table}

Table~\ref{tab:fano_comparison} summarizes the different sensitivities
of the three Fano factors to heating along the equilibrating edges.
For the conventional auto-correlation, the heating correction grows as
$\mathcal T^{-1/2}$ as the transmission is reduced and can therefore
dominate over the intrinsic QPC contribution even in the
weak-transmission regime, $\mathcal{T}\ll 1$. The cross-correlation is considerably more
robust: its heating correction scales as $\mathcal T^{1/2}$ and hence
vanishes for $\mathcal T\rightarrow0$. The clean Fano factor further
removes the dominant outer-arm heating contribution, leaving only a
higher-order correction of $O(\mathcal T^{3/2})$. Thus, $F_{\rm clean}$ provides a more reliable way to extract
$F_{\rm QPC}$, and hence the quasiparticle charge $e^*$, than
$F_{\rm auto}$ and $F_{\rm cross}$. 

\begin{figure}[t]
 \centering
\includegraphics[width=1.0\columnwidth]{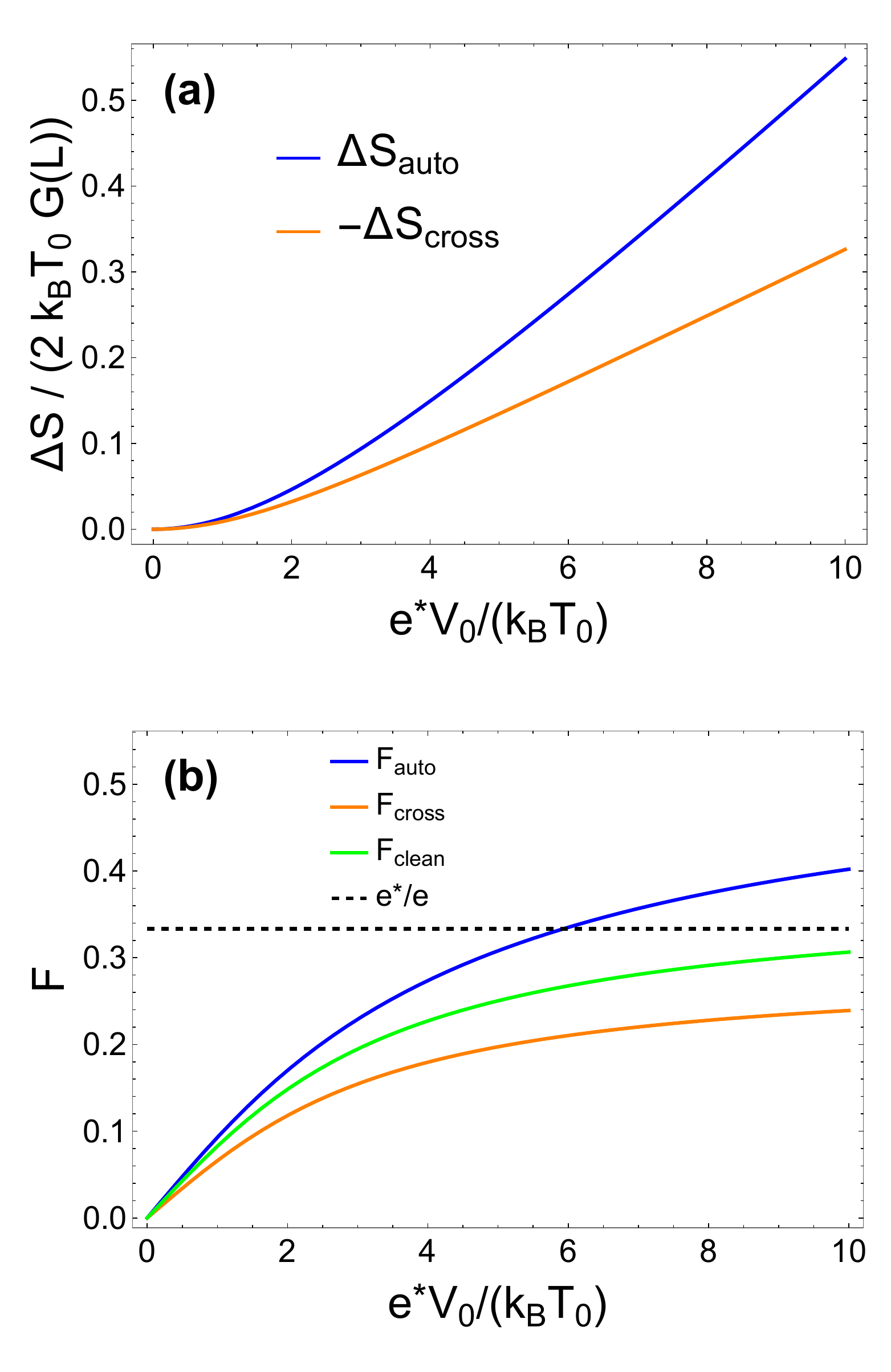}
\caption{
Bias dependence of the excess current noises and the corresponding Fano
factors, calculated from the full expressions without expanding in
$\mathcal T$. The calculations are performed in the strong
charge-equilibration regime, $\ell_{\rm ch}\ll L$. 
(a) Excess auto-correlation $\Delta S_{\rm auto}$ and negative
cross-correlation $-\Delta S_{\rm cross}$, normalized by the
equilibrium Johnson--Nyquist noise
$2k_BT_0G(L)$, as functions of the dimensionless bias
$e^*V_0/(k_BT_0)$, where $G(L) \equiv 2 G_0/3$ is the two terminal conductance of the fully equilibrated $\nu = 2/3$ edge. 
(b) Fano factors $F_{\rm auto}$, $F_{\rm cross}$, and $F_{\rm clean}$
as functions of the same dimensionless bias as in (a).
The horizontal dashed line indicates the quasiparticle charge
$e^*/e=1/3$. This plot shows that $F_{\rm clean}$ provides a more reliable extraction of the quasiparticle charge than $F_{
\rm auto
}$ and $F_{\rm cross}$ for FQH edges with upstream modes. 
For the plots, we use $L=3\ell_{\rm th}$, and $\mathcal{T} = 0.1$. 
}
 \label{fig:Fanofactor}
 \end{figure}

Figure~\ref{fig:Fanofactor} shows the bias dependence of the excess noises
obtained from the full expressions without expanding in $\mathcal T$.
As shown in Fig.~\ref{fig:Fanofactor}\textcolor{blue}{(a)}, both $\Delta S_{\rm auto}$ and
$-\Delta S_{\rm cross}$ increase with bias, but the auto-correlation
grows more rapidly. Their difference,
$\Delta S_{\rm auto}+\Delta S_{\rm cross}$ is precisely the excess
outer-arm noise as shown in Eq.~\eqref{eq:arm_noise_isolation}. The growing separation between the two curves thus reflects the
bias-enhanced heating contribution associated with the outer-arm
noise-generating spots.

The resulting Fano factors are shown in Fig.~\ref{fig:Fanofactor}\textcolor{blue}{(b)}.
The auto-correlation Fano factor $F_{\rm auto}$ substantially overestimates the quasiparticle
charge as the bias is increased, owing to the heating contribution
from the outer-arm spots. The cross-correlation Fano factor $F_{\rm cross}$  is considerably less
sensitive to this heating, but remains below $e^*/e$ at finite
transmission because of the residual outer-arm heating contribution
and the non-negligible factor $1/(1+\mathcal T)^2$ multiplying the
intrinsic QPC noise contribution, see Eqs.~\eqref{eq:noise_auto} and
\eqref{eq:noise_cross}. However, our newly introduced $F_{\rm clean}$ remains much
closer to the desired value $e^*/e$ with increasing bias than
$F_{\rm auto}$ and $F_{\rm cross}$. This numerical result, obtained
without a weak-$\mathcal T$ expansion, confirms the analytical hierarchy
summarized in Table~\ref{tab:fano_comparison} and demonstrates that
$F_{\rm clean}$ provides a more reliable extraction of the
quasiparticle charge in the presence of edge heating associated with
upstream modes.

We conclude this section by comparing our results with the phenomenological description proposed in Ref.~\cite{De_HOM_2025}. There, the effect of heating is incorporated through a single bias-dependent effective temperature, $T_e=\sqrt{T_0^2+\lambda^2(eV_0)^2/k_B^2}$, characterized by a phenomenological parameter $\lambda$. Within our framework, by contrast, the temperatures entering the different noise contributions are determined independently by the energy transport along the edge. In particular, the intrinsic QPC partition noise in Eq.~\eqref{eq:qpc_noise} is governed by the incoming temperature $T_{\rm in}$, whereas the equilibration-induced noise generated along the arms is controlled by $T_{\rm out}$, see Eqs.~\eqref{eq:excess_auto} and \eqref{eq:excess_cross}. Both temperatures depend on the propagation length and on the degree of thermal equilibration. Consequently, the heating contribution cannot in general be represented by a single effective temperature common to the QPC and the arms. Moreover, our formulation shows explicitly how the spatially distinct noise-generating regions contribute differently to the auto- and cross-correlation noises.

\section{Conclusion and outlook}
\label{sec:Conclusion}
We have developed a theoretical framework for analyzing shot noise in FQH edges with counterpropagating modes,
using the equilibrated $\nu=2/3$ edge as a representative example.
Starting from a line-junction model, we obtained the
voltage and temperature profiles generated by charge and thermal
equilibration and incorporated the corresponding local current
fluctuations into a four-terminal QPC geometry. This approach allows us
to distinguish the noise generated along the edges from the intrinsic
partition noise generated at the QPC.

A central result of our analysis is that the auto- and
cross-correlations respond very differently to heating along the
equilibrating edges. In the strong charge-equilibration and
weak-transmission regime, $\mathcal{T} \ll 1$, the heating correction to the conventional
auto-correlation Fano factor scales as $\mathcal T^{-1/2}$, whereas that
to the cross-correlation scales as $\mathcal T^{1/2}$. The
cross-correlation is therefore considerably less sensitive to edge
heating in the weak-transmission limit. We further showed that the
combination
$\Delta S_{\rm auto}+\Delta S_{\rm cross}$ directly isolates the excess
noise associated with the outer-arm noise-generating spots. Using this
relation, we constructed a new Fano factor $F_{\rm clean}$, which removes
the dominant outer-arm heating contribution using only experimentally
accessible quantities. Its remaining heating correction scales as
$\mathcal T^{3/2}$, allowing a more reliable extraction of the
quasiparticle charge at small transmission.

More generally, the framework developed here is not restricted to the
$\nu=2/3$ edge. For complex edges with several modes, co-propagating
modes that equilibrate rapidly can be grouped into effective sectors
characterized by a common local voltage and temperature and by their
total conductance. The decomposition of the measured drain-current
correlations into contributions from local fluctuation sources, as
exemplified by Eqs.~\eqref{eq:noise_auto} and
\eqref{eq:noise_cross}, can then be generalized to such multi-mode
edges. For a given edge structure, one needs to determine the effective
charge and heat transport equations, the corresponding voltage and
temperature profiles, and the local noise generated by the relevant
equilibration and QPC-partitioning processes. These quantities can then
be incorporated into the same framework for noises developed
here. This opens a route to analyzing more complex FQH edges, including
candidate non-Abelian states at $\nu=5/2$, where upstream charged or
neutral modes can similarly generate heating and complicate the
extraction of quasiparticle charges from shot noise measurements.

\begin{acknowledgments}
We acknowledge funding from the Swedish Vetenskapsr\r{a}det via Project No.~2023-04043 (C.S.), from the Deutsche Forschungsgemeinschaft (DFG, German Research Foundation)--- Project No.~559260877 (J.P.), and from Korea NRF---Grant No. RS-2026-25492880 (J.P.).
\end{acknowledgments}

\bibliography{Refs}

\end{document}